\documentclass[sigconf]{acmart}

\usepackage{subcaption}
\usepackage{multirow}

\AtBeginDocument{%
  \providecommand\BibTeX{{%
    \normalfont B\kern-0.5em{\scshape i\kern-0.25em b}\kern-0.8em\TeX}}}

\renewcommand{\paragraph}[1]{\vspace*{6pt} \noindent \textbf{#1.}\;}
\newcommand{\sparagraph}[1]{\hfill\\[4pt] \noindent \textbf{#1.}\;}

\newcommand\x{\par\hangindent1em\makebox[1em][l]{$\bullet$}}
\newcommand{\update}[1]{\textbf{}\textcolor{black}{#1}}

\copyrightyear{2025}
\acmYear{2025}
\setcopyright{acmlicensed}\acmConference[CHI '25]{CHI Conference on Human Factors in Computing Systems}{April 26-May 1, 2025}{Yokohama, Japan}
\acmBooktitle{CHI Conference on Human Factors in Computing Systems (CHI '25), April 26-May 1, 2025, Yokohama, Japan}
\acmDOI{10.1145/3706598.3713086}
\acmISBN{979-8-4007-1394-1/25/04}

\begin{document}

\title
[Exploring Data-Driven Advocacy in Home Health\update{c}are Work]
{Exploring Data-Driven Advocacy in Home Health\update{c}are Work} 


\author{Joy Ming}
\email{jming@infosci.cornell.edu}
\orcid{0000-0002-4289-5012}
\affiliation{
    \institution{Cornell University}
    \city{Ithaca}
    \state{New York}
    \country{USA}
}

\author{Hawi Tolera}
\email{hht25@cornell.edu}
\affiliation{
    \institution{Cornell University}
    \city{Ithaca}
    \state{New York}
    \country{USA}
}

\author{Jiamin Tu}
\email{jt794@cornell.edu}
\affiliation{
    \institution{Cornell University}
    \city{Ithaca}
    \state{New York}
    \country{USA}
}

\author{Ella Yitzhaki}
\email{ella.yitz@gmail.com}
\affiliation{
    \institution{Cornell University}
    \city{Ithaca}
    \state{New York}
    \country{USA}
}

\author{Chit Sum Eunice Ngai}
\email{cn336@cornell.edu}
\orcid{0000-0001-5192-7445}
\affiliation{
    \institution{Cornell University}
    \city{Ithaca}
    \state{New York}
    \country{USA}
}

\author{Madeline Sterling}
\email{mrs9012@med.cornell.edu}
\orcid{0000-0002-9928-4407}
\affiliation{
    \institution{Weill Cornell Medicine}
    \city{New York}
    \state{New York}
    \country{USA}
}

\author{Ariel Avgar}
\email{aca27@cornell.edu}
\orcid{0000-0002-2656-0060}
\affiliation{
    \institution{Cornell University}
    \city{Ithaca}
    \state{New York}
    \country{USA}
}

\author{Aditya Vashistha}
\email{adityav@cornell.edu}
\orcid{0000-0001-5693-3326}
\affiliation{
    \institution{Cornell University}
    \city{Ithaca}
    \state{New York}
    \country{USA}
}

\author{Nicola Dell}
\email{nixdell@cornell.edu}
\orcid{0000-0002-6119-705X}
\affiliation{
    \institution{Cornell Tech}
    \city{New York}
    \state{New York}
    \country{USA}
}

\renewcommand{\shortauthors}{Ming et al.}

\begin{abstract}

This paper explores opportunities and challenges for data-driven advocacy to support home care workers, an often overlooked group of low-wage, frontline health workers. First, we investigate what data to collect and how to collect it in ways that preserve privacy and avoid burdening workers. Second, we examine how workers and advocates could use collected data to strengthen individual and collective advocacy efforts. Our qualitative study with 11 workers and 15 advocates highlights tensions between workers' desires for individual and immediate benefits and advocates' preferences to prioritize more collective and long-term benefits. We also uncover discrepancies between participants' expectations for how data might transform advocacy and their on-the-ground experiences collecting and using real data. Finally, we discuss future directions for data-driven worker advocacy, including combining different kinds of data to ameliorate challenges, leveraging advocates as data stewards, and accounting for workers' and organizations' heterogeneous goals.



\end{abstract}



\maketitle

\section{Introduction}
Data-driven advocacy contends with many potential benefits and burdens. While it allows minoritized groups to collect evidence to address inequities, it also exposes them to potential surveillance and other data-related harms.
Research has shown how low-wage, frontline workers can use data to hold employers accountable, effectively turning the apparatus of surveillance on those who are usually the observers---a ``reversal of the Foucauldian panopticon'' \cite{Kelly_Garrett2006-oc}. For example, gig workers have subverted workplace algorithmic monitoring by collecting data about their working conditions \cite{Toxtli2021-mv} and systems like  Turkopticon \cite{Irani2013-gz} and Shipt Calculator \cite{Calacci2022-lh} have helped reduce information asymmetry and build worker power. On the other hand, critics point to the challenges of collecting data without adding responsibilities or risks for already precarious and overburdened workers \cite{Silberman2016-ci}. Moreover, data-driven advocacy may use data in ways that obscure worker voice, elevate statistics over people \cite{Spektor2024-mo}, or conflict with workers' values \cite{Pierre2021-gb}.


In this paper, we investigate how home care workers and advocates could collect and use data to improve working conditions without incurring potential harms. The duality of potential benefits and burdens of data-driven advocacy is especially poignant for home care work because data has the power to highlight the concerns and contributions of this invisibilized workforce \cite{Hatton2017-wr}, but also the potential to expose sensitive information that could harm vulnerable workers and clients \cite{Mateescu2021-sc}. Home care workers provide professional care services to older adults and people with disabilities in their homes, helping in activities of daily living, giving social support, and monitoring clinical symptoms. Despite their crucial role in promoting the well-being of vulnerable clients, these workers' labor is not always legally protected \cite{Kennedy2015-ja} and they face challenging working conditions and financial precarity \cite{Stacey2011-ax}. Data-driven advocacy holds potential in this context because, while home care work has been traditionally considered ``non-technical'' or involving little technology \cite{Dombrowski2017-ju}, it is becoming increasingly datafied. In fact, home care workers are now often compared to platform workers \cite{Poon2023-lr}, as their work is frequently shaped by digital tools and data-driven management systems. Additionally, technology has proven pivotal in organizing and connecting this spatially isolated workforce \cite{Poon2021-xb}.

Towards realizing data-driven advocacy for home care workers, this paper contributes a qualitative study that explores the
following research questions:
\begin{enumerate}
    \item[\textbf{RQ1:}] What data should be collected, and how, to address the needs of workers and advocates?
    \item[\textbf{RQ2:}] How could workers and advocates use the data collected for data-driven advocacy?
\end{enumerate}
To answer these questions, we conducted a multi-phase study with 11 workers and 15 worker advocates to design and pilot a data-driven advocacy intervention.
To design our intervention, we started with the open-source WeClock smartphone application that collects data on work and wages in a ``privacy-preserving way'' \cite{noauthor_2022-sd}. Based on feedback from advocate ideation sessions and iterative design with workers, we adapted 
the application to the needs of home care work. We then piloted the intervention and collected data from workers through a two-month field deployment. Finally, we generated insights based on this data and discussed these insights with workers and advocates.


Our analysis found trade-offs between data comprehensiveness, privacy, and work, as well as how the data could be used for individual, local, and policy advocacy. With regards to data collection (RQ1), we found that the workers' attitudes impacted their interactions with the intervention, including what they felt was important to report or not. Additionally, the workers' attitudes diverged from prior literature and the advocate concerns, such as being more open to sharing their data with their employers and more cautious about their clients' data. Finally, we saw how worker attitudes evolved through interactions with the intervention, becoming more open to using new technologies. 

With regards to data uses (RQ2), we found that while the individual workers got a lot out of reflecting on their working conditions, the real power of the data was at the aggregate level. The data insights from the pilot confirmed the advocates' knowledge of what workers experienced and they discussed different future directions for using the data, from broadening its scope to making it more specific to existing advocacy campaigns.

Our findings raise two tensions in data-driven advocacy based on worker and advocate attitudes. We saw instances where workers' and advocates' attitudes conflicted, as individual workers wanted immediate benefits to allay the burden of data collection and advocates prioritized more collective and long-term benefits. Their attitudes also conflicted with the realities of data collection---they had high expectations for how data could transform advocacy but noted limitations on organization resources and persuasiveness. Based on these tensions, we suggest future data-driven advocacy build power using a combination of numbers to attract attention and stories to hold it, leverage the positions of advocates to alleviate burdens on workers, and tailor efforts around specific organizational goals. 

By exploring data-driven advocacy in home health care work, our paper exposes the complex nuances of this context and illuminates how these may inform other instances of data-driven advocacy with low-wage, frontline workers, especially for workers who are minoritized, isolated, or engaged in care work. Ultimately, our research seeks improved recognition of the crucial contributions home care workers make and to promote the well-being of the workers, the clients they support, and all who benefit from care. 

\section{Related Work}

\subsection{Home Care \update{Work in the United States} } 

Due to demand from an increasingly aging population and a high rate of turnover in a challenged workforce, there will be an expected 5.5 million job openings in the \update{United States} home care workforce by 2032 \cite{PHI2024-yk}. 
However, current \update{labor} policies do not always adequately protect these workers, who are disproportionately immigrants and women of color. Home care workers were not covered by the Fair Labor Standards Act \update{which established minimum wage laws in 1938} until a loophole was resolved \update{decades later} in 2015 \cite{Kennedy2015-ja}. \update{Even when new laws are passed, such as the updated Public Health Law \S 3614-f passed in 2022 in New York State that promised increases in minimum hourly wage, misappropriated funds meant that home care workers were not able to benefit} 
\cite{New_York_State_Department_of_Labor2022-ch}. 
\update{Currently, i}n New York State, \update{outdated policy still does not}
extend basic worker protections to home care workers\update{. For example, overnight shifts that extend to 24-hour workdays for home care workers still frequently occur in New York City, an issue} No More 24 \cite{No_More_242024-or} \update{is trying to rectify through Department of Labor claims and city policy change like the No More 24 Act (Int 175). Other recent policy campaigns aim to continue to increase minimum wage for home care workers through the Fair Pay for Home Care campaign} \cite{noauthor_2022-jy} \update{or establish secure funding for home and community-based services through the Care Can't Wait campaign \cite{noauthor_2022-pr}}.

There are many challenges to advocating for home care worker voice, including that their situations vary by region \cite{Mareschal2006-qn} and that they are located between public and private realms \cite{Gruberg2017-me}. This raises important questions around how the workers are classified (i.e., as independent contractors or employees) \cite{Boris2006-vx}, the need for targeted education and training \cite{Margolies2008-vd}, or the role of cooperative organizations in supporting workers~\cite{Pinto2021-zq}. Moreover, a commonly raised concern is that home care workers are ``fragmented'' or ``atomized''---challenging to identify because they are geographically dispersed and often do not congregate at a central location like a traditional factory or office building \cite{Rhee2009-ra}.

While home care has traditionally been considered ``non-technical'' like many other low-wage jobs \cite{Dombrowski2017-ju}, it is increasingly shaped by digital tools and data-driven management systems \cite{Poon2023-lr,Kuo22}. This shift means that while
workers are increasingly burdened by new technologies that influence their work, they also have to rely on digital tools 
to advocate for better working conditions~\cite{Sterling2020-ix,Tseng2020-mi}.
Much of the prior HCI research has focused on the design of novel, top-down technological interventions, including voice assistants \cite{Bartle2022-nm,bartle2023machine}, internet of things \cite{Tian2023-tf}, 
machine learning \cite{Gappa2022-gi}, or virtual reality \cite{Reppou2024-gi}. Other research focuses on how technology-driven interventions often fall short of their intended goals by failing to consider the needs of workers. This line of work broadly examines the sociotechnical characteristics of home care, long-term care, and caregiving. For example, scholars have examined the broken promises of home care robots \cite{Wright2023-pf}, surveillance of electronic visit verification systems \cite{Mateescu2021-sc}, the increase of data work for long-term care workers \cite{Sun2023-ot}, and the technological burden of caregiving \cite{Chen2013-yq}. 

A few prior studies have tried to bridge the gap between technology interventions and advocacy for home care workers.  \citet{Poon2023-lr, Poon2021-xb} created computer-mediated peer support programs for home care workers to help them navigate difficult workplace conditions, while \citet{Ming2024-vg} explored the potential for smartphone-based systems to record and report instances of wage theft. These examples highlight the potential of technology to unite the voices of these workers, who often have limited opportunities to connect with each other due to the lack of shared physical workspaces. We now describe the body of work at the intersection of technology and advocacy in more detail. 



\subsection{Data-Driven Advocacy}
\label{rw-advocacy}

Data has long been used to advocate for and build collective power among low-wage, minoritized workers.
Starting from community-based self-surveys \cite{Broom1952-bq} to algorithmic audits \cite{Vecchione2021-vc}, minoritized groups have used data to highlight forces of oppression. Scholars have termed this  ``\textit{sousveillance}'' \cite{Mann2004-cp}, meaning ``\textit{to watch from below},'' a play on top-down surveillance.
In the framework of \citet{Darian2023-ai}'s functions of data in social good, data-driven worker advocacy demonstrates the potential for data to be used as an ``amplifier'' of worker voice \cite{Wilkinson2020-ag}, an ``activator'' to build solidarity among workers (e.g., crowd worker collective action \cite{Salehi2015-jw}), an ``incubator'' of new ways to substantiate worker issues and reduce information asymmetry (e.g., employer ratings \cite{Irani2013-gz}, wages and earnings \cite{Calacci2022-lh}), and a ``legitimizer'' of these issues (e.g., exposing working conditions \cite{Raval2020-ey}). 
However, the literature on data-driven worker advocacy also highlights several challenges, including how well data captures working conditions and how doing so could lead to privacy harms or additional work for workers.

\paragraph{Quantifying Work}
Starting from time-motion studies in industrial scientific management \cite{Khovanskaya2019-cm}, both employers and activists have tried to capture working conditions with metrics. Many of these efforts for a ``Quantified Workplace'' \cite{Adler2022-tx} are driven by employers in the hopes of increasing productivity, which often conversely results in more worker burnout and poor working conditions because of the data collection itself \cite{Underman2024-mi}. Moreover, these measures often do not categorize and measure intangible components of work 
that could be considered ``invisible work'' \cite{Crain2016-xi}, ``shadow work'' \cite{Illich1981-pe}, or ``ghost work'' \cite{Gray2019-ai}. This is especially relevant to care work, where ``care algorithms'' \cite{Wright2023-pf} attempt to itemize care tasks but fail to capture all contributions the workers make. Research has looked specifically into how to operationalize care work, including focusing on the costs of care work through scales for burnout or emotional exhaustion \cite{Wharton2009-cg} or mapping activities of care workers through extensive sensors and activity logs \cite{Nafus2016-lt}. Our research builds on these previous measures by exploring methods that are ``data-plus'' \cite{Spektor2024-mo}, in that they include both numbers and stories to paint a more holistic picture of work. 

\paragraph{Legitimacy or Surveillance}
Having data about working conditions could not only enhance the ``legitimacy'' of the work, but also enact ``surveillance'' that restricts worker autonomy \cite{Star1999-xo}. Following \citet{Citron2021-vo}'s framework for privacy harms, this surveillance could also lead to 
economic harms if employers retaliate \cite{Ming2024-vg}, discrimination harms with ableist assumptions of what care looks like \cite{Oduro2021-dl}, or relationship harms as vulnerable clients are effectively surveilled along with the workers \cite{Mateescu2021-sc}. \citet{Nissenbaum2009-fr}'s framework of contextual integrity also notes the importance of questions around the privacy and public divides, which is especially relevant to home care work because it occurs in private homes but involves public health.
Building on these insights, our work explored participatory privacy approaches \cite{Mir2021-xo} to solicit opinions ``from below'' early in the design process \cite{Zong2023-ot}.

\paragraph{Worker Burden}
Additionally, participatory privacy likely requires extra efforts on the part of the workers---as does the whole enterprise of data-driven worker advocacy. As \citet{Silberman2016-ci} note, any instances of workers holding employers accountable require the workers to put in the extra work of organizing. These additional issues could result in stalling and friction among workers as they move towards collective action \cite{Salehi2015-jw}. As home care workers already face serious hurdles in learning how to use technology \cite{Tseng2020-mi}, any technology or data-driven solution would require more learning. Finding the energy to overcome this ``technostress'' \cite{Varanasi2021-xm} is especially challenging for workers that ``have a lot going on in their lives'' \cite{Ming2024-vg}. Research in social movement theory has identified a few different approaches that either reduce burdens for workers or increase incentives for them to participate \cite{Coleman1966-rm}. Our research \update{seeks to} carefully minimize the technological burden on workers by designing tools that amplify collective action and leverage the efforts of other stakeholders in the care ecosystem.

\subsection{Our Contribution}
Much of the work on data-driven advocacy has primarily focused on gig workers, leaving the potential benefits and challenges of such advocacy largely unexplored in other contexts, like home care work. Our research builds on the existing literature to explore different approaches to navigating the challenges around data comprehensiveness, privacy, and burden. We focus on the essential and increasingly datafied context of home care to explore 
approaches to data-driven advocacy for these frontline workers. 




\section{Methods}

\begin{figure*}
    \centering
    \includegraphics[width=0.7\linewidth]{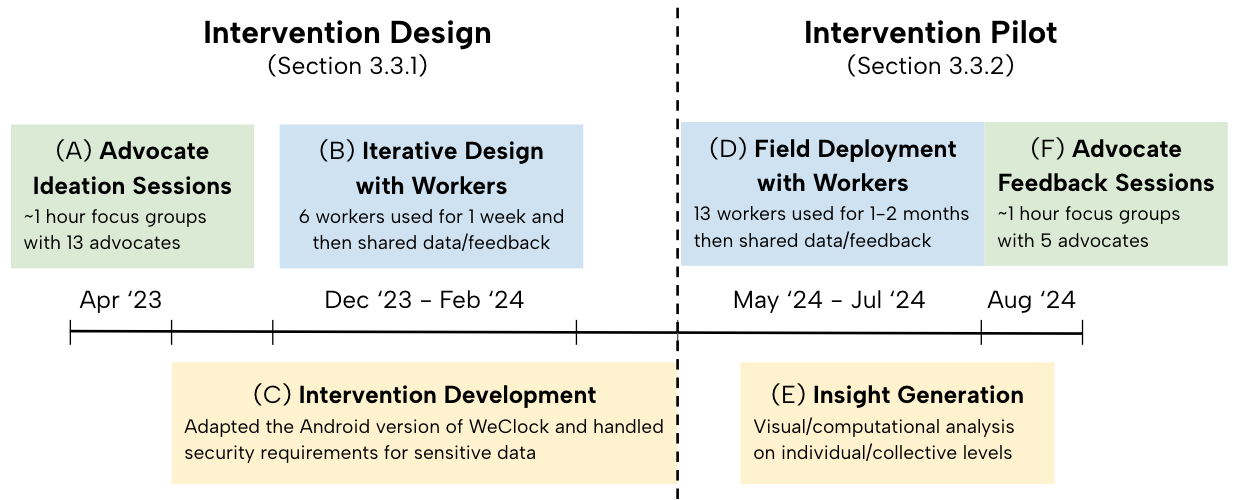}
    \caption{Timeline view of study activities, including the intervention design (detailed in Section~\ref{sec:intervention-design}) and pilot (detailed in Section~\ref{sec:field-study}) phases, with advocate sessions (green), worker trials (blue), and coding/computation (yellow).}
    \Description[Timeline view of study activities with descriptions of each of the activities: Ideation, Intervention Development, Iterative Design, Field Deployment, Insight Generation, Feedback]{A timeline view of the different activities that are part of the study. The activities are split into two parts by a dotted line and each of boxes is color coded and includes a short description of each phase. In the Design phase we have three boxes: a green box indicating advocate session for Ideation (Apr'23), yellow box indicating coding/computation for Intervention Development (Apr'23-May'24), and blue box indicating worker trial for Iterative Design (Dec'23-Feb'24). In the Pilot phase, we have three boxes: a blue box for Field Deployment (May'24-Jul'24), a yellow box for Insight Generation (Jun'24-Aug'24), and a green box Feedback (Aug'24).}
    \label{fig:timeline}
\end{figure*}

We conducted a multi-phase, IRB-approved study to explore the potential benefits and challenges of data-driven worker advocacy in the home care context. First, we adapted an open source smartphone tool to specifically fit the data goals and workflows of the home care context (Section~\ref{sec:intervention-design}), engaging with 11 advocates in ideation sessions and 6 workers through iterative design in a week-long trial. We then deployed the intervention over a month with 13 workers to evaluate how well it could address the challenges of data-driven worker advocacy (Section~\ref{sec:field-study}). 
Finally, the insights from this data collection were presented to four worker advocates. 
Figure~\ref{fig:timeline} summarizes our study phases.

\subsection{Research Partners}

Our research involved workers and advocates that were based in the State of New York in the United States of America.
Our main partner, Healthcare Workers Rising (HWR) \cite{noauthor_2022-dd} is a nonprofit organization based in Upstate New York that focuses on improving working conditions for home care workers through free trainings, monthly meetings, and policy advocacy. The organization is also affiliated with 1199SEIU, the largest healthcare union in the US and are involved in efforts to organize these workers. Our secondary partner, New York Caring Majority (NYCM) \cite{noauthor_2022-vi}, is a movement of seniors, people with disabilities, family caregivers, and domestic and home care workers from across New York State. They are primarily focused on advocacy through specific policy initiatives and interacting with legislators.

\subsection{Participants}

\update{Our study engaged two main types of participants: advocates and workers. We wanted to leverage the expertise of different stakeholders, including the individual lived experiences of the workers and the collective experiences the advocates had of engaging with many workers and legislators. Moreover, we included heavier involvement from advocates in order to balance the epistemic burden placed on the marginalized workers.}

We recruited advocate participants through our partners. We aimed to solicit opinions of advocates affiliated with both worker-centered Healthcare Workers Rising (HWR) and policy-oriented New York Caring Majority (NYCM). 
In total, our 
advocate participants included 9 advocates affiliated with HWR and 5 affiliated with NYCM (see Table~\ref{tab:advocates}).

We recruited workers mainly through HWR.
Workers needed to have
at least one consistent client and use an Android phone. A total of 14 workers used the application (see Table~\ref{tab:demographics}), six of whom also participated in the initial short trial. There were three participants who used the intervention but were unavailable to return to participate in a feedback interview. All of the workers were women, Black or African American, and US citizens by birth. They were a mix of the two types of home care workers: personal care aides (PCAs) who are focused on more social aspects of health and home health aides (HHAs) who are focused on more medical aspects. 

\begin{table*}
\caption{Advocate self-reported demographic information}
\label{tab:advocates}
\footnotesize
\begin{tabular}{llllll}
\toprule

PID & Phase(s)           & Affiliation & Gender    & Position                               & Years of Experience \\
\midrule
A1  & A \& F & HWR         & Woman     & New Organizing Coordinator / Director & 15 years            \\
A2  & A             & HWR         & Man       & Organizer                              & 19 years            \\
A3  & A             & HWR         & Man       & Organizer                              & Unknown                   \\
A4  & A             & HWR         & Woman     & Organizer                              & 7 years             \\
A5  & A             & HWR         & Man       & Organizer                              & 7 years             \\
A6  & A \& F & HWR         & Woman     & Organizer                              & 5 years             \\
A7  & A             & HWR         & Man       & Organizer                              & 19 years            \\
A8  & A             & NYCM        & Nonbinary & Advocacy Specialist                    & 1 year              \\
A9  & A             & NYCM        & Woman     & Advocacy Specialist                    & 3 years             \\
A10 & A             & HWR         & Man       & Organizing Director                    & 18 years            \\
A11 & A             & HWR         & Man       & Senior Communications Coordinator      & Unknown                   \\
A12 & A             & NYCM        & Woman     & NY Political Director                  & 29 years            \\
A13 & A \& F & NYCM        & Woman     & Lead NY State Organizer                & 16 years            \\
A14 & A             & HWR         & Nonbinary & Organizer                              & 6 years             \\
A15 & F             & NYCM        & Woman     & Statewide Systems Advocate             & 1 year   \\          
\bottomrule

\end{tabular}
\end{table*}

\begin{table*}
    \caption{Worker self-reported demographic information}
    \label{tab:demographics}
    \footnotesize
    \begin{tabular}{lllllll}
    \toprule
    PID & Phase(s) & Position  & Gender    & Race/Ethnicity    & Education Level   & Years of Exp. \\
    \midrule
    W1 & B \& D & HHA & Woman & Black or African American & High school graduate & 5 \\
    W2 & B \& D  & PCA, Other & Woman & Black or African American & Some college & 7 \\
    W3 & B & PCA & Woman & Black or African American & High school graduate & 20 \\
    W4 & B \& D  & PCA & Woman & Black or African American & 2-year degree & 1 \\
    W5 & B \& D  & PCA & Woman & Black or African American & High school graduate & 10 \\
    W6 & B \& D  & Unknown & Unknown & Unknown & Unknown & Unknown \\
    W7 & D & HHA & Woman & Black or African American & Some college & 6 \\
    W8 & D  & PCA & Woman & Black or African American & Some college & 8 \\
    W9 & D  & HHA, PCA & Woman & Black or African American & Some college & 4 \\
    W10 & D  & PCA & Woman & Black or African American & Some college & 4 \\
    W11 & D  & PCA & Woman & Black or African American & Some college & 7 \\
    \bottomrule
    \end{tabular}
\end{table*}

\subsection{Study Procedure}


\subsubsection{Intervention Design}
\label{sec:intervention-design}

\begin{figure*}[]
    \caption{Screenshots showing functions of the intervention. \label{fig:screenshots}}
    \begin{subfigure}[b]{0.22\textwidth}
        \caption{Overview of modules}
        \includegraphics[width=\textwidth]{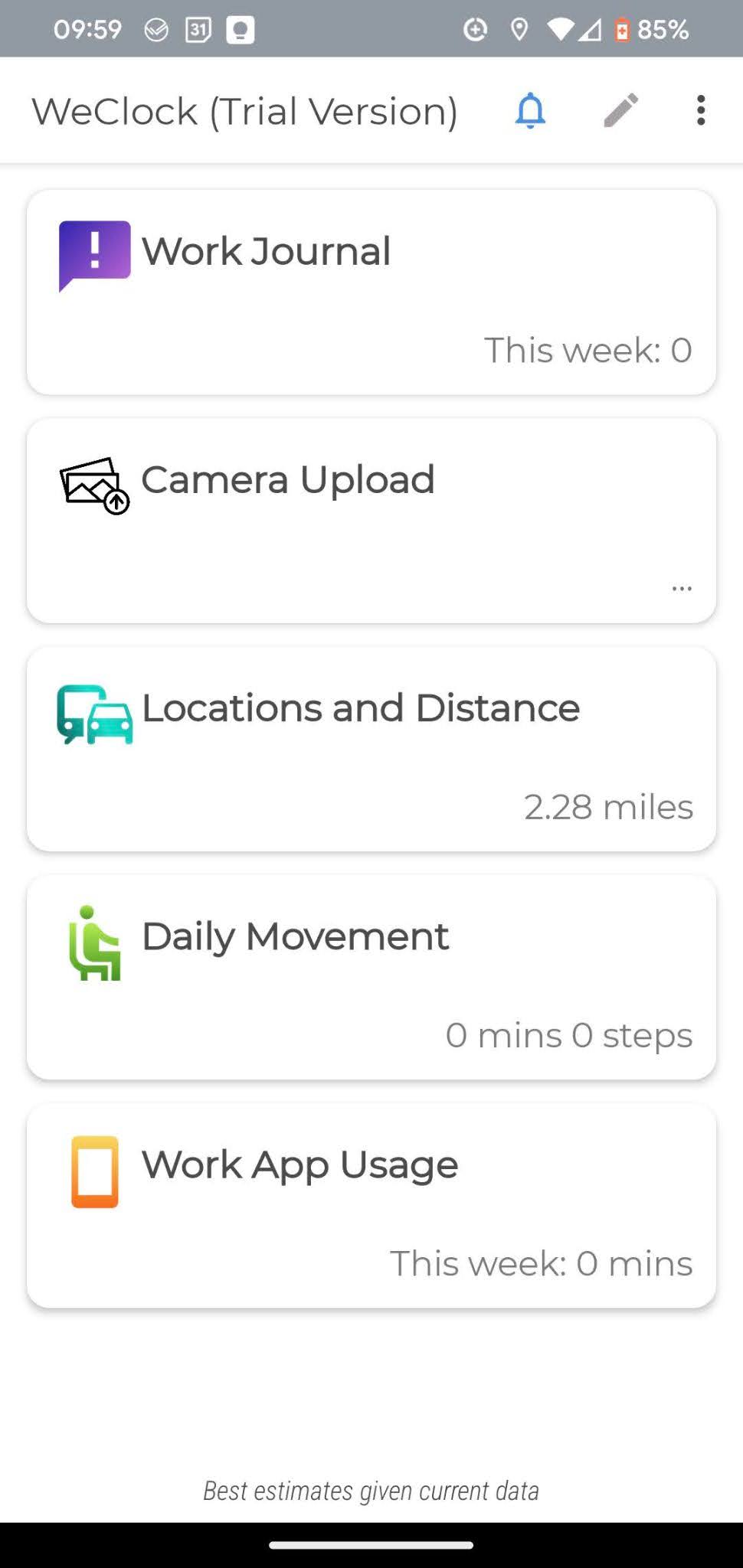} 
        \label{fig:screenshot-overview} 
    \end{subfigure}
    \begin{subfigure}[b]{0.22\textwidth}
        \caption{Open-text journal}
        \includegraphics[width=\textwidth]{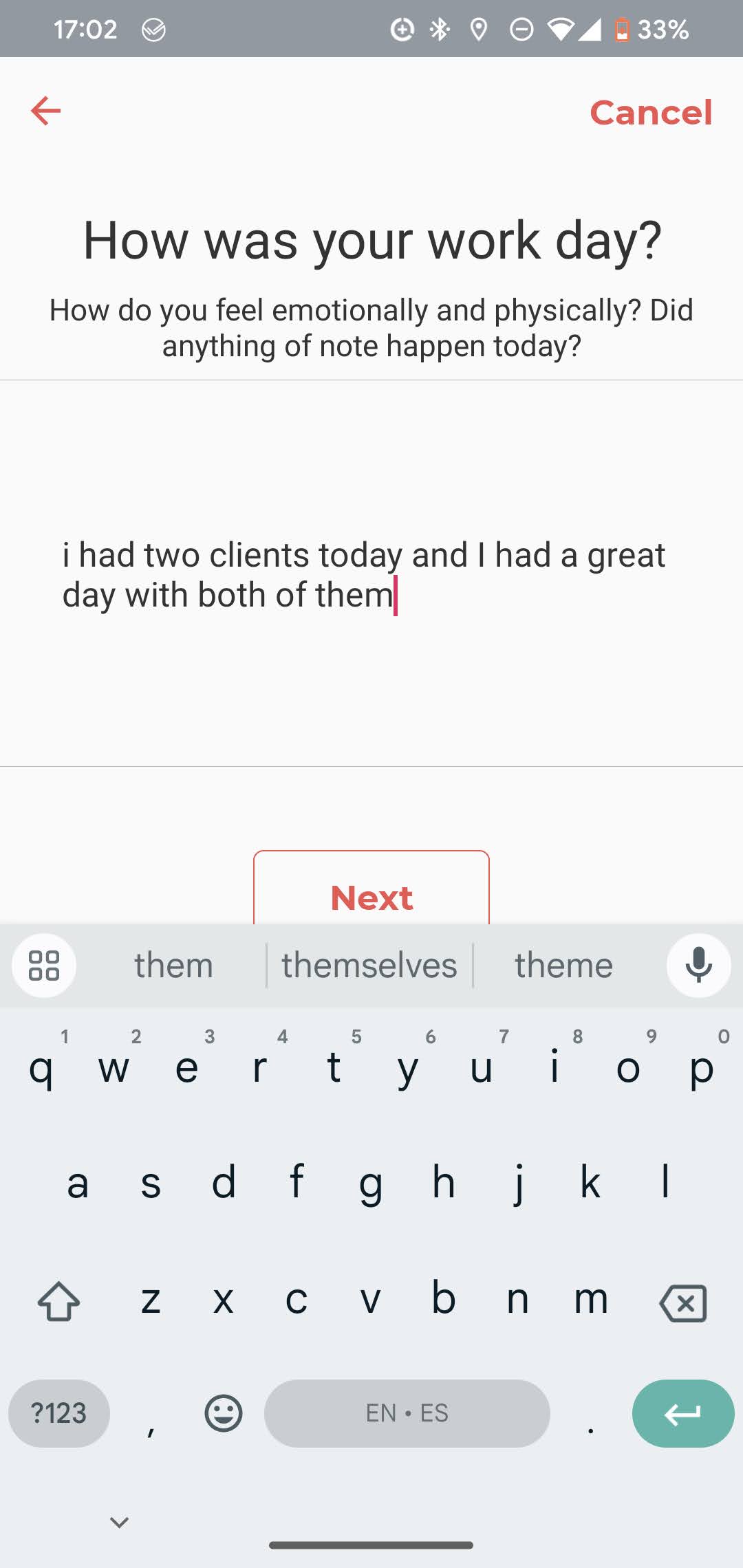}
        \label{fig:screenshot-open} 
    \end{subfigure}
    \begin{subfigure}[b]{0.22\textwidth}
        \caption{Multiple-choice journal}
        \includegraphics[width=\textwidth]{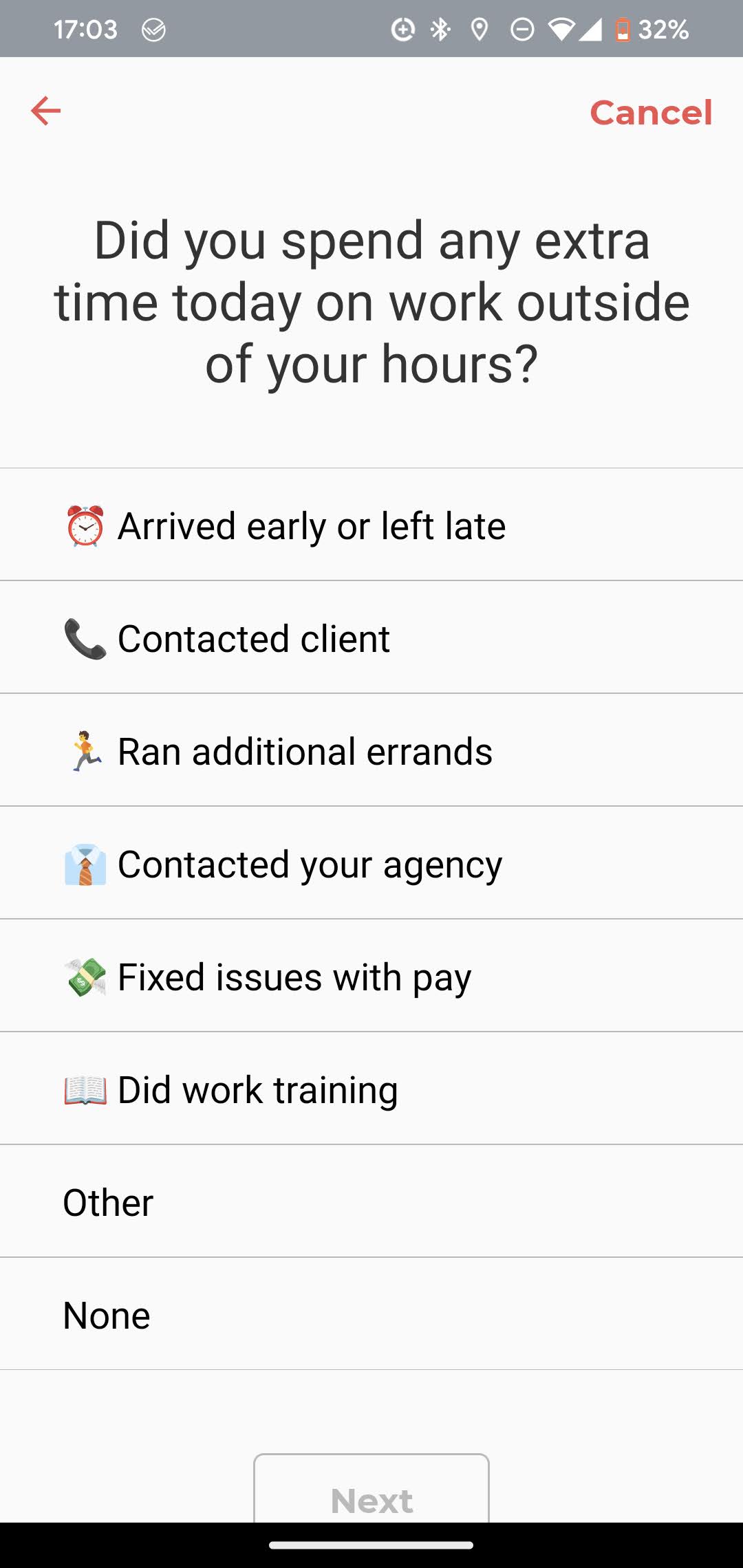} 
        \label{fig:screenshot-choice} 
    \end{subfigure}
    \begin{subfigure}[b]{0.22\textwidth}
        \caption{Location tracking}
        \includegraphics[width=\textwidth]{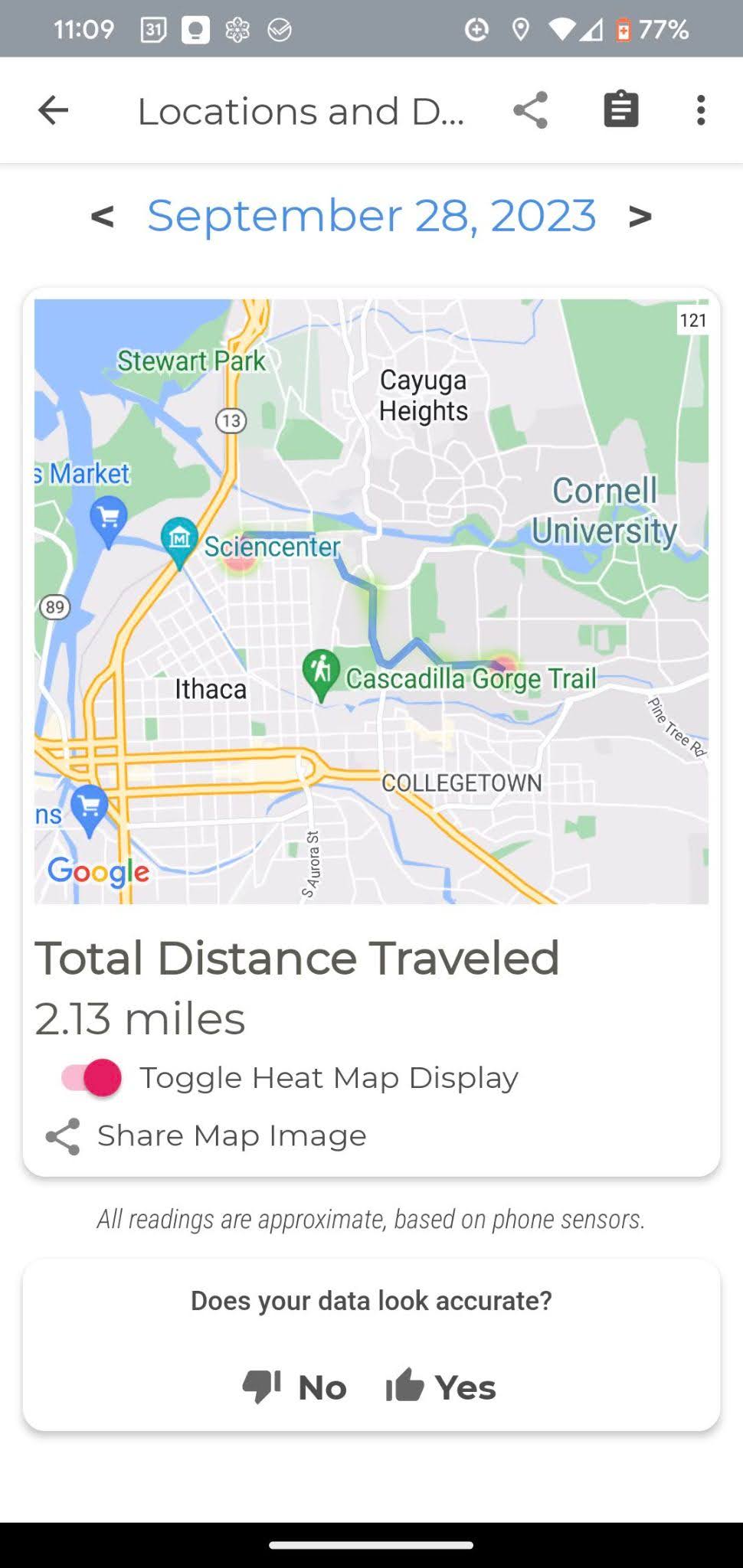}
        \label{fig:screenshot-location} 
    \end{subfigure}
    \Description[Four different screenshots from the WeClock smartphone application.]{The first screenshot shows an overview of the modules in the WeClock app, including ``Work Journal,'' ``Camera Upload,'' ``Locations and Distance,'' ``Daily Movement,'' and ``Work App Usage.'' The second screenshot shows a stage of the journaling, which asks the question ``How do you feel today'' and gives the options of ``Exhausted,'' ``A bit exhausted,'' ``Not exhausted or energized,'' ``A bit energized,'' and ``Energized'' with corresponding emojis. The third screenshot shows the question "How was your work day? How do you feel emotionally and physically? Did anything of note happen today?" with a response "i had two clients today and i had a great day with both of them." The fourth screenshot shows a map with the distance traveled marked and ``Total Distance Traveled'' for ``September 28, 2023.'' }
\end{figure*}

We started with open-source smartphone application WeClock to design our intervention. The application helps workers collect data on their working conditions, through user-inputted observations about their workdays and background tracking of location, movement, and work app usage \cite{noauthor_2022-sd}\update{.} Figure~\ref{fig:screenshots} \update{presents} sample screenshots \update{of the intervention, including an overview of the types of tracking functionality supported by the application (Figure~\ref{fig:screenshot-overview}), samples of prompts for both open-text and multiple-choice journal entries (Figures~\ref{fig:screenshot-open}, \ref{fig:screenshot-choice}), and the output of one day of location tracking (Figure~\ref{fig:screenshot-location})}. We selected this application because of its ability to collect data on both emotions-based and systems-based invisible work~\cite{Ming2023-cq} (i.e., stress/burnout tracking and time tracking), while also minimizing the workload through the use of phone sensors to automatically collect data. The application also accounts for privacy (i.e., stores data locally and gives workers control over data sharing), a key consideration raised in prior research \cite{Ming2024-vg}. Moreover, its open-source nature means that the application already exists and could be disseminated to other low-wage workers.

\begin{figure}[]
    \includegraphics[width=\linewidth]{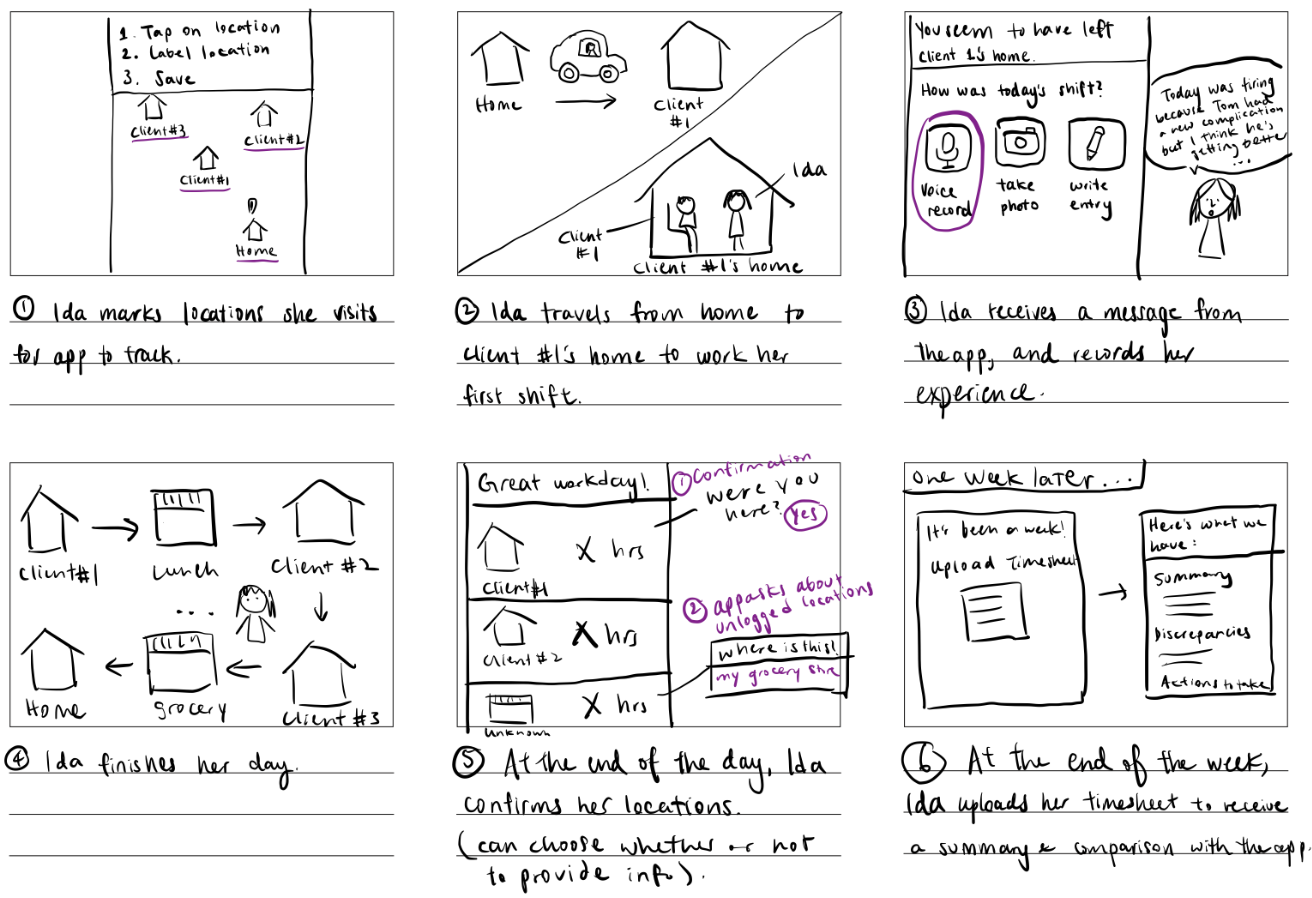}
    \caption{Storyboard depicting the proposed data collection flow presented in the advocate focus groups.}
    \Description[Six panels describing a home care worker Ida's use of a sample intervention.]{In the first panel, there is an image of an where Ida marks locations she visits like her home and clients' homes. In the second panel, there is an image showing Idea traveling from her home to client \#1's home to work her first shift. The third panel shows idea receiving a message from the app to record her experience at her shift. In the fourth panel, Ida travels between her clients homes, lunch, and the grocery before returning home to finish her day. in the fifth panel, the app shows idea confirming her locations for the day and choosing whether or not to provide the information. Finally, the sixth panel shows Ida uploading her timesheet and paystub one week later to receive a summary of her wages and hours by the app.}
    \label{fig:storyboard} 
\end{figure}

\paragraph{\update{(A)} Advocate Ideation Sessions}
We held five initial focus groups with \update{13} worker advocates to explore how data could be useful in their advocacy efforts and what concerns they had about data collection methods. Focus groups lasted approximately 60 minutes, were conducted over Zoom, and were recorded and transcribed with the participants' consent. During the focus groups, participants were asked to brainstorm what data about home care workers might be helpful from the perspective of advocates, workers, and policymakers. Finally, we discussed potential workflows, including collecting worker data using a smartphone application. We first considered the user journey using a storyboard \update{that depicted a worker marking locations, traveling between clients, and reporting on her daily experience} (see Figure~\ref{fig:storyboard}) and gathered concrete feedback on the application features using screenshots.

\paragraph{\update{(B)} Iterative Design with Workers}
We then conducted initial user trials, in which we asked \update{6} workers to use the application for a week. Workers were asked to fill out a form that included informed consent and a demographic survey. The consent form was re-explained and participants were given an opportunity to ask questions during the first 30-45 minute Zoom session where they downloaded and set-up the application. Workers then used the application for one week, including writing a work journal for every day they worked. Finally, workers attended a 30-45 minute Zoom session where they uploaded their data through a secure portal and gave feedback on the application. Workers received a USD \$30 \update{gift card} for participating in both sessions and sharing a week of data. \update{The main iterations we made throughout this process were around the types of questions we asked about their workday (e.g., consolidating the separate questions about physical and emotional energy/exhaustion into one question) and how workers uploaded their data (e.g., reducing the number of steps to download the data and upload it to the secure portal).}

\paragraph{\update{(C)} Intervention Development}
Based on \update{the advocates and workers' feedback}, we \update{adapted some of the} features in the WeClock application \update{to be more relevant to the home care context}. \update{We developed some changes to the intervention based on advocate feedback, iterated on the features with worker feedback, and then finalized the changes before the intervention pilot.}
\update{Ultimately, the main changes to the application were that we} \update{gave home-care-specific examples in structured questions about the context and impacts of unpaid work based on advocate and worker discussions}, created an option to upload paystubs and timesheets to contextualize their location data \update{based on existing worker workflows}, and provided an option for automatic data upload that workers could opt into \update{due to worker demand}. 

\subsubsection{Intervention Pilot}
\label{sec:field-study}


We collected two types of data with the intervention: narrative and numeric. For the narrative data, we were inspired by the experience sampling method \cite{Fisher2012-ma} or ecological momentary assessment \cite{Shiffman2008-zm} frameworks to take measurements 
of exhaustion, job satisfaction, and control as proxies for quantifying burnout from emotional labor \cite{Wharton2009-cg} and instances of invisible work reported by previous studies, such as \cite{Ming2023-cq}. In our journal entries, we asked workers to rate their energy/exhaustion on a five-point scale, share information about their day in open text, and use prepopulated multiple choice fields to report 
invisible work. For the numeric data, we collected GPS location data 
and records of schedules/paystubs to  determine how long workers spent in their work location compared to the hours they were paid, giving us numbers on invisible work in the form of unpaid time. 

\begin{figure}[]
    \centering
    \includegraphics[width=\linewidth]{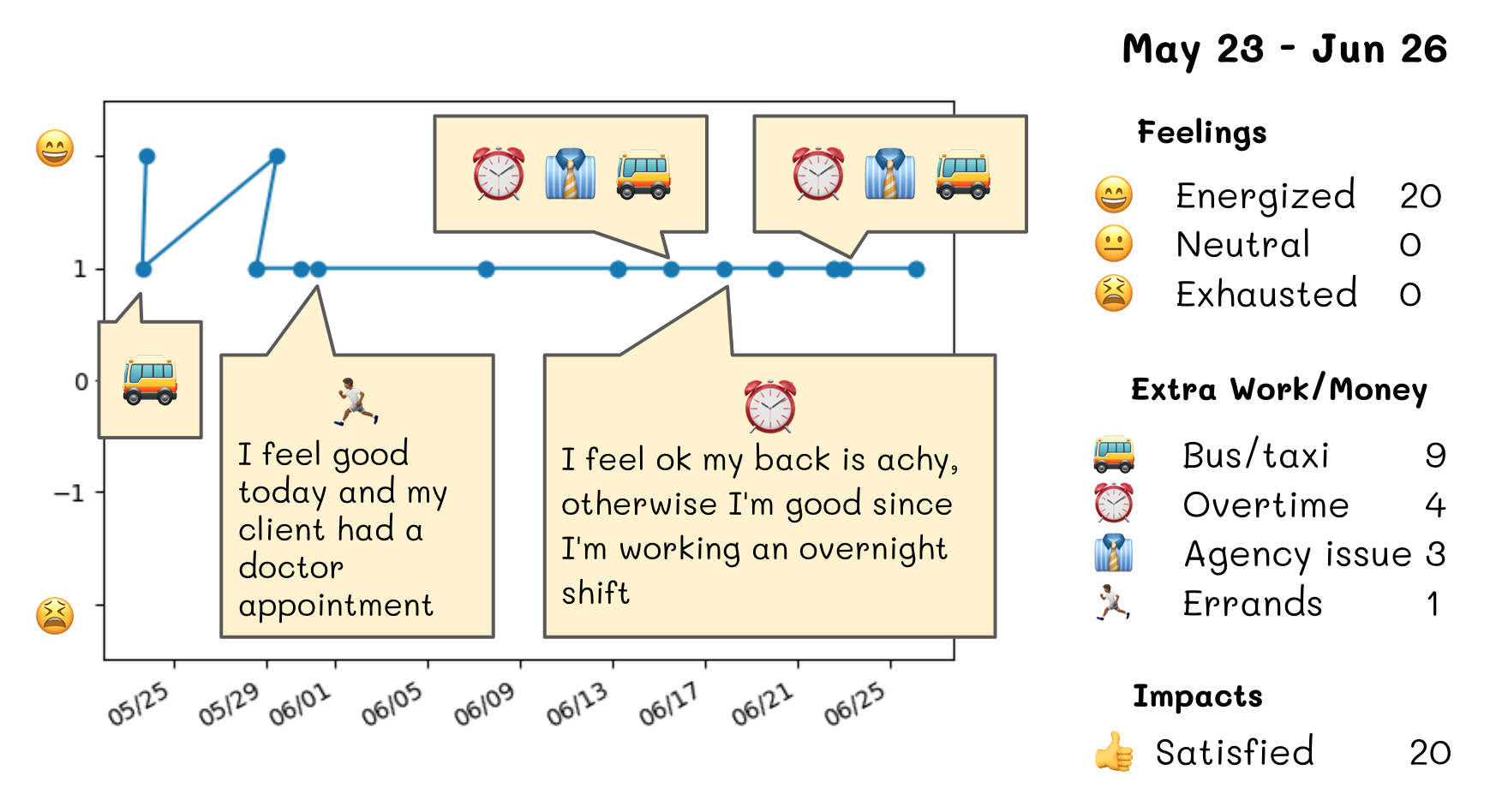}
    \caption{\update{Summary visualization of an} individual\update{'s} journal entries \update{throughout the study period}}
    \label{fig:individual}
    \Description[A visual representation of the journal entries of a given worker for the period of work.]{There are two sections to the visualization. On the left, there is a chart where the x axis is date and the y axis is feelings (from lowest exhausted to highest energized). The workers' ratings of their feelings show up as a line graph, with points mainly marked as "energized" and two points of "very energized." Some of these points include more context from the journal entries, including icons like bus/taxi or alarm clock and a snipped from the journal entry. On the right hand side, there is a title that says the date range and counts for each of the categories of feelings, extra work/money, and impacts.}
\end{figure}

\paragraph{\update{(D)} Field Deployment with Workers} 
\update{13 workers completed all stages of the field deployment.}
At the start of the deployment, 
the workers downloaded the application, after which we walked them through an onboarding questionnaire, completed a sample journal entry, uploaded sample hour tracking, and set up either manual or automatic data uploading. 
Workers that did not already participate in the iterative design phase were assisted by a representative from our partner organization and those that had participated updated their application remotely. Participants received a \$25 USD giftcard for completing onboarding. 
Next, the workers then used the application for four to eight weeks. 
\update{During this time, we periodically checked if any of the participants were missing data for a longer period of time (i.e., missing data upload for 3+ days or missing journal entries for 1+ week) and followed up with the participant over text or phone.}
Participants receiv\update{ed} a \$10 USD giftcard for each week of data shared.
Finally, the workers attended a feedback session over Zoom where we walked them through insights generated from their uploaded data. Participants were asked about their experiences using the application, their reactions to visualizations of their data (e.g., Figure~\ref{fig:individual} \update{illustrates an overview of one worker's journal entries throughout the study period, including snippets of the open text that give context to one of their structured answers}), and their perspectives on data collection and sharing. Participants received a \$25 USD giftcard for attending this final feedback session.

\paragraph{\update{(E)} Insight Generation}
Based on the data workers collected, we generated insights about individual workers and the group of workers as a whole. We analyzed the periodic measures of GPS information of the location data using mobility analysis tools (i.e., \texttt{scikit-mobility} \cite{Pappalardo2022-zc}) to generate insights about overtime and commute time for each worker. On an aggregate level, we used visualization libraries (i.e., 
\texttt{seaborn} \cite{Waskom2021-ay}, 
\texttt{geopandas} \cite{Jordahl2014-zx}) to understand workers' schedules, including overlaps in shifts, average work hours, and common locations (see Figure~\ref{fig:visualizations} for examples and Section~\ref{sec:collective} for further discussion). The journal data included structured multiple choice and open-ended questions about each worker's experience, which we analyzed using aggregate counts, simple correlation (i.e., Pearson correlation using \texttt{statsmodels} \cite{Seabold2010-uw}), and basic natural language processing techniques (i.e., topic modeling using \texttt{tomotopy} \cite{Unknown2024-yq} and \texttt{jsLDA} \cite{Mimno2018-yy}) (see Figure~\ref{fig:counts} and Section~\ref{sec:invisible-work} for further discussion). These analyses and visualizations were presented for feedback.

\paragraph{\update{(F)} Advocate Feedback Sessions}
We wanted to understand which types of data collected would be most useful in and explore what it would take to realize successful data-driven advocacy. We held four 60-90 minute Zoom sessions 
with \update{a total of 5} advocates, 
including three who were part of early study phases and two who were new due to staffing changes at the partner organizations. 
During the sessions, advocates were presented with insights based on workers' location and journal data. We asked which visualizations, if any, might be useful for advocacy, including how the data aligns with existing or future advocacy campaigns. Advocates also discussed their perspectives on the tradeoffs involved in realizing data-driven worker advocacy, including best case scenarios and minimum requirements.

\subsection{Data Analysis}

Our study data included recordings of Zoom focus groups, training, and feedback sessions, transcribed via \texttt{Whisper} \cite{Radford2023-xy} and \texttt{PyAnnote} \cite{Bredin2023-eh} libraries. We also analyzed other interactions with workers, including text messages and notes from phone calls during the data collection period. 
We used thematic analysis to analyze our data \cite{Braun2006-kv}. First, we used open coding with gerunds \cite{Charmaz2011-po} to develop a codebook with 70 codes (i.e., ``\textit{pointing out systemic issues about caregiver shortage},'' ``\textit{identifying employer violations}, ``\textit{describing feelings towards extra work}'') across the different phases. We had two to three researchers code the transcripts separately and meet to discuss and reconcile codes. Then, we organized the data into themes (i.e., ``\textit{misunderstanding of technology},'' ``\textit{logistical challenges}'') that represented our study findings.

\subsection{Positionality}

Our team is part of a larger coalition researching home care work from multidisciplinary perspectives, including public health, labor relations, health policy, and information science. In addition to research, team members have also participated in broader advocacy activities. We have partnered with HWR for over two years, studying questions related to home care and technology. 
As representatives of a large research institution, we are aware of the need to navigate power dynamics as we try to center the voices of workers and advocates in an equity-driven, collaborative approach.

\section{Findings}

\update{We organize the findings based on research question and, within each, present workers' and advocates' initial \textbf{expectations} from the intervention design phase (methods described in Section~\ref{sec:intervention-design}) and their resulting \textbf{experiences} from the intervention pilot phase (methods described in Section~\ref{sec:field-study}).}
First, we discuss findings relevant to collecting data for data-driven advocacy (RQ1), organized around the key considerations of data comprehensiveness, privacy, and burden (Section~\ref{sec:data-collection}). We describe the decisions we made during iterative design, responding to feedback advocates raised about these considerations in our ideation sessions. We also juxtapose advocates' expectations with their experiences of how these factors played out in the field deployment. Then, we discuss our findings relevant to the uses of data (RQ2), including the potential to amplify benefits in supporting individual workers, collective organizing, and policy advocacy (Section~\ref{sec:data-uses}). We present the expectations discussed by advocates in the ideation sessions and workers during iterative design in conjunction with their experiences on how well the field deployment was able to realize those goals. 
Table~\ref{tab:findings} summarizes our findings \update{with rows corresponding to subsubsections}. 

\begin{table*}[]
    \centering
    \footnotesize
    \begin{tabular}{p{0.2\linewidth} p{0.35\linewidth} p{0.35\linewidth}}
        \toprule
        & Expectations & Experiences \\
        \midrule
            \ref{sec:data-collection} \textbf{Data Collection} & &  \\
            \midrule
            \ref{sec:invisible-work} \textit{\update{Capturing Holistic Data about Work}}
            & 
                \x Qualitative reports could document instances of invisible work and context around them
                \x Quantitative measures could count instances of unpaid work (i.e., overtime, commute)
            & 
                \x Responses influenced by attitudes towards extra work
                \x Computational analysis of qualitative data created numbers that could be used
                \x Numeric and narrative data led to more power
            \\ \cmidrule(lr){1-3}
            \ref{sec:privacy} \textit{\update{Addressing Data} Privacy \update{Concerns}} 
            & 
                \x Employers could retaliate with intervention use
                \x Workers would want control over when, how, and with whom their data is shared
            & 
                \x Workers updated beliefs overtime, including how to use privacy paradigms or consider privacy
                \x Wanted to share data for advocacy, some even with employers to hold them accountable
                \x More concern over privacy of client information, especially around location
            \\ \cmidrule(lr){1-3}
            \ref{sec:burden} \textit{\update{Reducing} Worker Burden \update{in Data Collection}} 
            & 
                \x Resistance to learning/downloading a new technology
                \x Automated location tracking would simplify process
            & 
                \x Initial frustration, eventual confidence with tech
                \x Logistical difficulties led to inconsistency 
                \x Physical limitations of automatic technology led to more burden
            \\ 
        \midrule
            \ref{sec:data-uses} \textbf{Data Uses} & & \\
            \midrule
            \ref{sec:individual} \textit{\update{Empowering} Individual \update{Reflection and Improvement}} 
            & 
                \x More awareness of working conditions for self-care
                \x Identifying meant advocating for issues
            & 
                \x Great space to vent frustrations, reflect on extra work
                \x A few cases of wage issues
                \x Knowledge was not enough to change the situation
            \\ \cmidrule(lr){1-3}
            \ref{sec:collective} \textit{\update{Building Solidarity and} Collective\update{ly Organizing Workers}} 
            & 
                \x Connecting workers meant mobilizing towards action
                \x More data could build a better understanding of worker issues/lives across employers
            & 
                \x Statistics were helpful to confirm information
                \x Want more specificity for different work circumstances or different groups of workers
            \\ \cmidrule(lr){1-3}
            \ref{sec:policy} \textit{\update{Creating More Caring} Polic\update{ies}} 
            & 
                \x Harder statistics on worker realities
                \x Inform better distribution and pay for workers to address caregiver shortage
            & 
                \x Some advocates want larger scale intervention to make statistical claims for more workers
                \x Some advocates want data to be more tightly tied to their existing campaigns/concerns
            \\
        \bottomrule
            
    \end{tabular}
    \caption{Summary of findings}
    \label{tab:findings}
\end{table*}

\subsection{Data Collection: Balancing Comprehensive Data With Privacy and Burden}
\label{sec:data-collection}

Prior literature noted three key considerations in collecting data from workers to advance advocacy: (1) how well the intervention captures holistic data about work (i.e., numeric and narrative aspects, less tangible aspects \update{\cite{Wright2023-pf, Spektor2024-mo}}); (2) how well the intervention protects privacy of worker data, especially with location tracking \update{(e.g., \cite{Mateescu2021-sc, Citron2021-vo})}; and (3) how well the intervention manages the added burden of data collection for workers \update{(e.g., \cite{Silberman2016-ci, Tseng2020-mi})} (see Section~\ref{rw-advocacy}). 

Through iterative design and field deployments, we found that a combination of numeric and narrative data helped contextualize the more invisible aspects of work, which could increase the power of the data for advocacy. We also found that workers did not conceptualize data privacy in the ways that the 
advocates predicted---the workers were concerned with the privacy of their client's data, but were willing to share their own data with their employers. Finally, regarding strategies to balance the responsibility of data collection with other worker priorities, we found that although workers were able to overcome their initial hesitation about technology, challenges in logistics and technological infrastructure were more difficult to overcome.

\subsubsection{Capturing Holistic Data About Work}
\label{sec:invisible-work}

\sparagraph{Expectations}
We aimed to collect both narrative and numeric data about home care workers' work, especially focusing on difficult to operationalize and otherwise invisible aspects of their work. As A11 noted, there is a lot of ``\textit{gray area}'' in home care work because ``\textit{sometimes people are asked to do all sorts of wild things that are far outside of their duties}.'' The advocates wanted ways to investigate this ``\textit{unpaid labor}'' (A15) and collect information about out-of-pocket costs workers are incurring, such as a worker spending time and money to help their client buy clothes for a wedding. 

\begin{figure}[t]
    \centering
    \includegraphics[width=\linewidth]{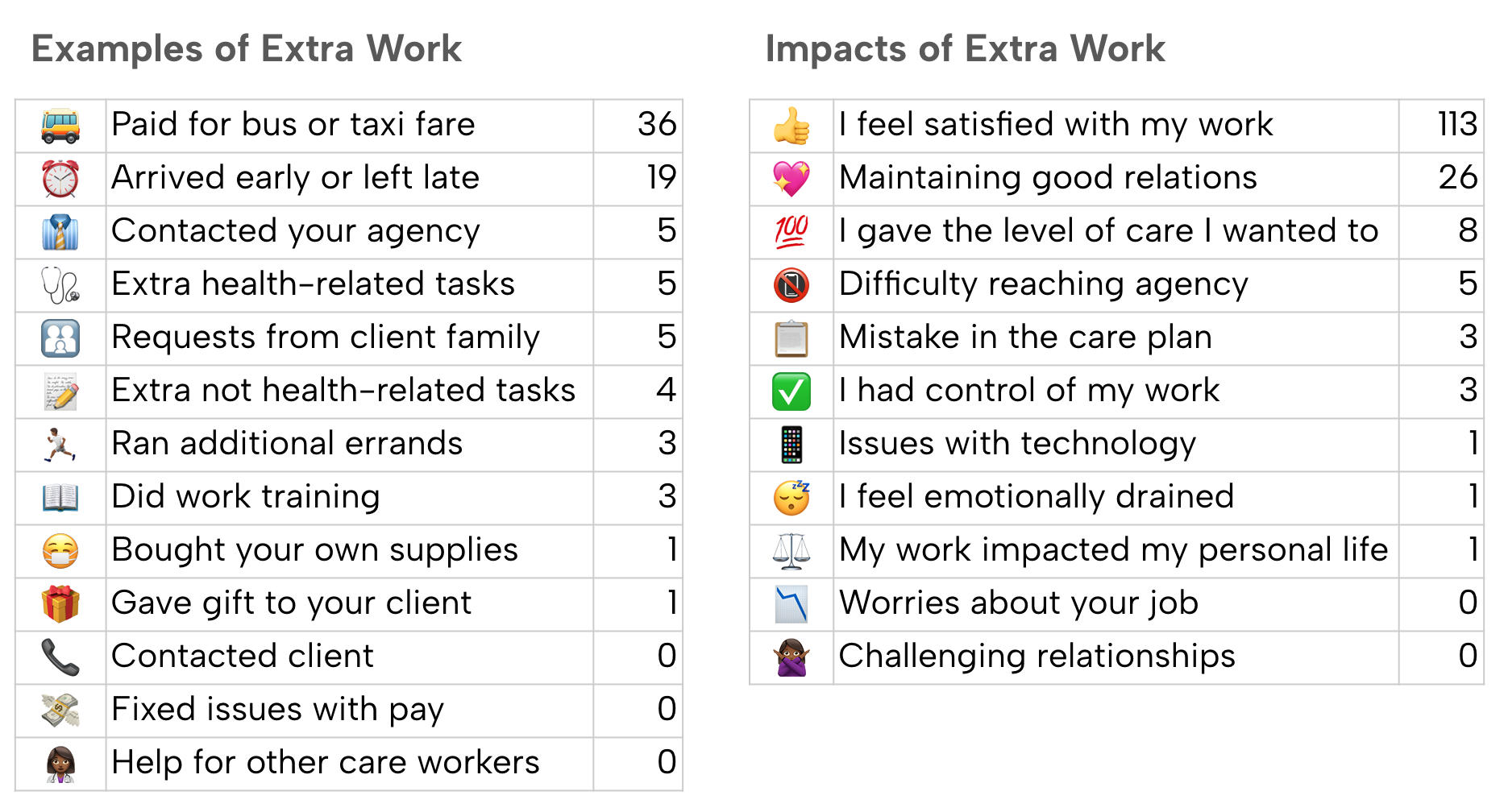}
    \caption{Counts of \update{examples and impacts} invisible work \update{reported in daily journal entries} across workers}
    \label{fig:counts}
    \Description[Tables that demonstrate the counts of examples of extra work and impacts of extra work.]{The figure consists of two tables: the table on the left displays examples of extra work mentioned in workers' journal entries along with their counts, while the table on the right shows the impacts of extra work, also with corresponding counts. Each row includes icons that visually represent the entries, such as a running emoji for the extra work of running additional errands. Extra work such as paying for commute and working longer hours have the biggest counts and seem to be very common amongst workers. Regarding the impact, many workers reported feeling satisfied with their work and noted that the extra efforts contributed to maintaining good relationships.}
\end{figure}

\paragraph{Experiences}
The qualitative and quantitative data did yield concrete examples of invisible work, but also raised tensions regarding what to collect and how it was collected or interpreted. On the qualitative side, we received a total of 138 journal entries, identifying individual and group examples of the different kinds of extra work and its impacts\update{.} Figure~\ref{fig:counts} \update{presents} an overview of these journal entries\update{, showing the counts for each of the examples and impacts of extra work the workers recorded within their journal entries during the study period}. 

Our qualitative data revealed that 
workers' reports depended on what they themselves perceived as ``extra'' or unpaid work. Many workers considered staying late with their clients or doing activities for their clients' family an integral part of providing care, but their 
attitudes towards this changed through the journaling process:
\begin{quote}
    ``\textit{The app actually got me thinking \dots about the little stuff that we do that we don't really get paid for \dots I think that being in this field, it's a given that you're going to end up going a little bit extra for the clients \dots 
    you're going to go that extra mile to make sure that they're okay, even if it's not part of your job.}'' (W7)
\end{quote}
Other workers did not even consider this ``\textit{little bit extra}'' as something worth reporting. W1 noted few examples of extra work because she felt like ``\textit{there wasn't nothing really extra---I just did everything I usually do}.'' The workers' attitudes towards extra work not only influenced their reports of what they were doing, but also how they felt about doing it. In most journal entries (113 out of 138), workers reported that they ``\textit{feel satisfied with [their] work}.''

We also encountered challenges combining quantitative and qualitative data. One was that the quantitative data alone was less effective at capturing invisible work. Quantitative data, in the form of GPS location, only showed where the worker was and for how long but lacked contextual information for what the worker was doing there and whether it was work-related or not. Moreover, we faced accuracy issues in the GPS measurements. Thus, we had to ask workers for more context about their GPS locations, which split their efforts away from providing high quality journal entries.

The advocates found the narratives very useful not just in how workers talked about their challenges, but to also see what were the recurring issues.
This sparked discussion around the power of numeric data in addition towards narrative data. A15 spoke about using a blend of numbers and narratives to capture invisible work holistically: ``\textit{big, hard statistical numbers can be helpful to attract attention \dots and people always want at least one in depth story about a person---I think we really need both}.''

\subsubsection{Addressing Data Privacy Concerns}
\label{sec:privacy}

\sparagraph{Expectations}
Prior literature \update{(e.g., \cite{Star1999-xo, Citron2021-vo, Mir2021-xo})} and ideation sessions raised data privacy as a concern in data-driven worker advocacy. Worker advocates like A1 noted that when other applications mandated workers to report their hours using GPS tracking, there was ``\textit{push back from caregivers, [saying] `I don't want my agency to know where I am}.' '' Additionally, in the ideation sessions, A11 expressed concerns about employer retaliation for people using the intervention, anticipating ``\textit{some sort of terms of employment clause prohibiting this kind of data collection}'' or resulting in fewer work opportunities to those who used the intervention. \update{For this intervention pilot, we wanted to try to collect the broadest amount of data to determine which would be the most useful for data-driven advocacy. One of our main goals was specifically to capture instances where workers were at work outside of their usual or predicted working hours, including cases where they have unexpected inconsistent or odd hours or if they had to stay late with a patient. D}ue to these \update{privacy} concerns, we started with a design that \update{gave workers control over their data. We allowed workers to save data locally to their phone so they have control over who gets to see it. Workers were also able to start and stop the sensors as they desired as well.}


\paragraph{Experiences}
However, the iterative design and field deployment phases demonstrated that workers' mental models of privacy differed from the expectations, particularly with regards to which data they wanted privacy protections for and how they believed privacy should be protected. For example, the specific privacy-protecting design patterns we used (i.e., storing data locally) conflicted with what the workers were expecting. W1 contacted us, saying she was ``\textit{reaching out to see if the app is working on y’all’s end; I wasn’t sure if [the data is] going through}.'' She assumed that the intervention would upload information to the researchers and advocates automatically. \update{Partially due to these mental models and to concerns around burden, after we developed the option to automatically upload the data, all of the workers opted for it, even though they had the option to manually select which parts of their data they wanted to send. This is likely because we were partnering with a worker advocacy organization they trust, rather than as independent researchers or with their employers.}

Generally, workers were open to sharing their data with certain people, especially if it would be used for advocacy. Many expressed optimism about how they felt their information might be utilized, including W5 who hoped the data could lead to improvements her working conditions:
\begin{quote}
    ``\textit{I think that if we're trying to improve working situations and conditions \dots I'm all for it. \dots We're not doing anything wrong. We're not doing anything illegal. We're not exposing our clients, so bring it on.}'' (W5)
\end{quote}
This was mostly in context of sharing their information with worker advocates, like W8, who noted that she would ``\textit{love [the  advocates] to see our information because they're the ones fighting for us to get help}.'' 

Rather than being completely opposed to sharing their data with their employers, as the advocates anticipated, the workers saw pros and cons for how this information could be used to hold employers accountable. W4 hoped that her information could be shared with her employers because it ``\textit{can make a difference \dots and they'll see the daily things we have to do and we have to go through}.'' Workers also raised different concerns to those raised by advocates, specifically in context of reporting invisible work. Workers felt that they might get penalized by their agencies for  
reporting the extra work, especially as their agencies have at times asked the workers to not get attached to their clients or do extra things for them. W8 expressed her fear that agencies could use journal entries to track ``\textit{some of the stuff we're not supposed to do}'' and that  ``\textit{saying what we're doing extra might get a person fired}.''

Workers also contrasted the original assumptions in that, rather than their own data, workers were more concerned about privacy of their client's information.
W7 said ``\textit{I don't care---people know who I am and what I share}'' but ``\textit{I just don't feel comfortable sharing a lot of my client's personal information}.''  
W1 shared concerns that the GPS location tracking might expose her client's address. She was the only participant who asked to turn off the location collection in the field deployment and in the feedback interview noted that ``\textit{if it was a building, it would be a little different because it wouldn't have their apartment number}.'' She also said that she would feel comfortable turning on the location tracking if it was collected at a lower resolution and if it were ``\textit{showing the area \dots as long as it's not showing the client's address}.''

Although most workers were not initially concerned about their privacy, 
several changed their perspective. 
Most started off feeling comfortable sharing their data as ``\textit{an open book}'' (W5). W4 started with a similar mindset, but, through more discussion about privacy, noted ``\textit{you do have me thinking},'' raising concerns about information such as address or phone number being leaked because, as she said: ``\textit{I know how people sell information \dots and then you get all these crazy phone calls}.'' She was also concerned because 
ultimately, ``\textit{someone can always have access}'' to her information.

\subsubsection{Reducing Worker Burden in Data Collection}
\label{sec:burden}

\sparagraph{Expectations}
We found that workers and advocates were cautious about new technological tools. 
A1 said her concerns stemmed from the fact that for HWR, ``\textit{a lot of our folks [(workers)] are older \dots [and] it seems like the older people get more hours too}.'' One of her concerns for ensuring the intervention collected consistent data was that ``\textit{people get frustrated with technology}.'' The advocates worried that 
workers would be more hesitant to adopt a new technology because of the steep learning curve for new tools. A15 said that workers would ``\textit{not love [the intervention] just because it's another thing on top of the other things that they're already having to do}.''  

The advocates also noted that workers might be hesitant to adopt a new technology due to their physical or built infrastructure. A1 raised concerns around the fact that ``\textit{people change phones so often---they break their phones, their phones are stolen, they're just not able to pay the bills}.'' They worried that with each change of phone or phone number, workers would be less likely to continue with the intervention. Additionally, advocates noted issues around phone storage space, which meant that workers might not want to download something new to their phones.

The advocates and workers emphasized the necessity of a tool that is straightforward and accessible for workers, ensuring it not only fits seamlessly into workers' routines but also gathers high-quality data without imposing significant burdens. Advocates in the ideation sessions said one of the main goals of the tool would be to ``\textit{simplify}'' (A7) their hour tracking and paystub checking process to be ``\textit{something that wasn't onerous}'' (A2). A2 also pointed out that ``\textit{the more you can automate it [(the process)], the better},'' with some advocates suggesting automation based on workers' schedules. 

Based on these concerns, 
we wanted the tool to operate primarily in the background, requiring minimal active engagement from workers, who would only need to use the intervention to fill out a work journal or upload timesheets and paystubs. This was in contrast to other time tracking applications that require workers to manually start and stop a timer to track their work hours. 

\paragraph{Experiences}
While both workers and advocates found the intervention useful and usable, we faced breakdowns in physical and logistical infrastructure that impacted data consistency and accuracy. While workers were initially hesitant about their ability to adopt a new technology during onboarding, most eventually felt comfortable with the intervention. At first, W3 expressed hesitation using the tool because ``\textit{it's my first time doing something like that}.'' 
In cases where workers felt frustrated with the technology, we leveraged tools like screensharing on Zoom and, in some cases, had a representative from HWR help workers in-person. 
Towards the end of the study, many workers felt more confident, including W3 who changed her mind to say: ``\textit{it was easy \dots [even though] I thought it was going to be hard}.''

During the data collection phase, workers also faced logistical and physical challenges. One logistical challenge was around workers' busy schedules and responsibilities. W5 noted her appreciation that the intervention fit well into her day-to-day because it ``\textit{naturally tracks the distance from my home to work \dots [and] I do the journal at the end of my shift}.'' However, she was not able to keep this up and she ``\textit{wasn't checking it every day because \dots things got a little crazy for a couple of weeks}.'' W6 also had issues remembering to use the intervention between her two different jobs. \update{Our team helped nudge workers towards engaging with the intervention. For example, we would remind workers to upload their timesheets or paystubs in the intervention every week, but only a few workers actually uploaded their timesheets and that was while they were directly being guided by the research team.}

Workers also faced challenges around physical infrastructure---%
two workers changed phone numbers during the study and three either broke or changed their phones. Nevertheless, they still shared their data with us and attended the feedback interview. However, some physical infrastructure challenges we\update{re} hard to overcome, such as problems with internet connectivity that led to imprecise location data. For example, we asked W6 for her weekly schedule to serve as ground truth and found that the data the intervention collected for W6's schedule was consistently delayed and shortened, potentially due to inconsistent connectivity.

The issues with physical infrastructure increased the burden on workers to manually supplement processes that were supposed to be automated. Often, we asked workers to verbally share their scheduled working times or confirm some of the data that we collected. Other times, workers themselves took on more tasks due to misunderstanding the technology. W7 said she was concerned about data accuracy and took the time to turn her location on and off when she was going to her other job so that location data from her other job ``\textit{doesn't interfere}.'' W5 was under the impression that the location tracking needed the intervention to be active and made sure to ``\textit{open up the app}'' before going to work.

\subsection{Data Uses: Supporting Individual, Collective, and Policy Advocacy}
\label{sec:data-uses}

After exploring how we could collect data to enable data-driven advocacy, we now examine how the data we collected could be used to amplify current advocacy efforts. 
At an individual level, we found that data helped workers reflect on their working conditions and identify issues. However, many felt that they  were not in a position to fully address these concerns on their own. On a 
collective level, advocates found that the data confirmed their existing priorities, but recognized that more effort was needed to encourage workers to connect and mobilize. Finally, advocates had policy goals to use the data to address more systemic problems around caregiver shortages, but felt they would either need larger scale data to make statistical claims or more specificity for targeted campaigns.

\subsubsection{Empowering Individual Reflection and Improvement}
\label{sec:individual}

\sparagraph{Expectations}
Both workers and advocates wanted workers to understand their work better in order to improve their individual working conditions. A8 noted the importance of this, telling us that they ``\textit{like that some of the pieces are not just collecting data, but actually being useful for the people using it as well}.'' A6 felt this type of data collection would help workers ``\textit{open their eyes to to things they probably never really pay attention to},'' such as how consistently staying late at their client's home affected their schedule and budget. Moreover, advocates and workers also noted the importance of the intervention as an opportunity for self-care. A5 said, ``\textit{a lot of caregivers will put a lot of other people and things before them}'' and this intervention could be an opportunity for workers to ``{\it talk about emotional regulation}'' (A11) and ``\textit{relieve their daily stress about their work}'' (A1). 

Another aspiration for the intervention was helping workers identify and resolve issues at work. Often, this took the shape of workers ensuring they were being paid for the hours they were working, helping workers ``\textit{know if they're being cheated or not when it comes to their paycheck}'' (A7). 
A2 felt this information could also be used to ensure employers were paying legally required sick time or other benefits. W10 wanted to keep track of her overtime and paid time off benefits because she did not trust her agency to give her what she was due. Moreover, workers thought the intervention could be especially helpful to track mileage to make sure they were reimbursed or to monitor wear and tear. 

\paragraph{Experiences}
Workers and advocates found the intervention useful in helping workers reflect on their work situation, but workers needed more support to actually change their conditions. The workers enjoyed the process of writing journal entries after each shift. W5 described the process of filling out the daily journal as ``\textit{romantic}''---an opportunity to talk about her day because she lives alone. W11 echoed the importance of journaling which allowed her to express things she cannot share with her family and friends without divulging private client information. 

Additionally, W10 found that filling out a journal entry at the end of each shift gave her an opportunity to think more about her own work and how it impacted her: ``\textit{when I answer some of the questions, I'll be thinking}.'' 
W8 framed this record of her work as ``\textit{self-recognition of what you're doing}.'' When she was reviewing her journal entries, she noted examples of how the data could help her learn about herself, including that ``\textit{I need to get more energized}'' or ``\textit{I can cut back on the gift giving a little bit}.'' She also described that this type of information might be helpful to bring to her doctor to describe how she was feeling. 

The intervention was able to identify some cases in which workers were not fully compensated for their work. W3 uploaded screenshots of her recorded hours and pay for the study period and, based on these documents, we found roughly 30 minutes of discrepancies between the hours our study tracked, the hours her employer tracked, and 
the hours she got paid on her pay stub. W9 noted that even the process of uploading her pay stubs to the intervention helped her catch that she was not being paid for her overtime on a holiday. 

However, some participants thought that having the data would not help, given the power dynamics in their workplace. W11 went into depth, saying that even though data on her working conditions ``\textit{would probably be interesting to know},'' she ultimately felt that ``\textit{it wouldn't make a difference to record it or not}.'' She, along with other workers, pointed out the difficulty of changing their behavior based on their data, such as finding more opportunities to rest or even coming up with specific actions based on their journal entries. 
More importantly, the workers felt they would not necessarily be able to advocate for themselves against their employers.
W6 felt she would not be able to make changes on her own: ``\textit{I can't defend myself \dots I would need somebody}.'' She wanted more support from other workers with similar issues or from HWR.

\subsubsection{Building Solidarity and Collectively Organizing Workers}
\label{sec:collective}

\sparagraph{Expectations}
Participants also noted the potential for the intervention to help workers connect with each other and build solidarity towards collective action, with the data ``\textit{showing workers that they're not alone in these issues}'' (A14). This idea comes from traditional organizing practices, starting from a ``\textit{validation of daily concerns}'' (A11) and then ``\textit{building power and moving people towards taking action and sharing in public their struggles}'' (A8). The advocates saw the potential for data about working conditions to encourage workers to connect with each other, saying it could ``\textit{give them that little extra push to say, `Damn, maybe I do need a union, maybe I do need to reach out to my coworkers'} '' (A6).


Moreover, advocates noted the potential for the data to help them better understand workers' realities to effectively advocate for them. 
Several advocates saw the benefits of ``\textit{trying to get the data for stuff we already know}'' (A12), whether in specific numbers or more detailed stories.
They mentioned that even though workers would come to them with concerns around wage theft, 
the advocates did not have an exact, actionable amount of how much money was lost in wages or out-of-pocket costs. 
A1 said that tracking how many times these types of issues occurred could help them advocate for institutionalizing protections and ``\textit{building contract language that addresses those issues}.'' 
A6 said this understanding would also be helpful for the advocates themselves to build connections with workers:
\begin{quote}
    ``\textit{It's more to understand what they're really going through, and to be sympathetic about how their job works for them, you know. And to really have a purpose of why we're organizing these people \dots it's not just because we need to organize them because it's our job, but it's the purpose behind it.}'' (A6)
\end{quote}

Advocates also saw the potential for the data to help reach more workers. A11 noted this was especially helpful in settings where they may not have many organizers to go around and connect with workers individually: 
\begin{quote}
    ``\textit{This type of information gathering tool not only is important for us to learn more about the workforce and what they're dealing with and how to advocate to resolve these issues, but also is a programming piece \dots to stay in touch with home care workers as we don't necessarily have the organizing infrastructure or resources [A1's team] has}.'' (A11)
\end{quote}

The advocates noted concerns about how to locate home care workers, an issue specific to this geographically dispersed workforce. As one advocate explained, ``\textit{we're not going shop to shop}'' (A11), referencing the traditional method of union building. There was no central agency where home care workers would go or with which the worker advocates could build a contract. The advocates noted that having data about where and when workers spend time could help them ``\textit{be somewhere where the most workers are gonna be}'' (A1). They mentioned wanting to match these locations to bus stops, areas of interest (e.g., laundromats), or living areas (e.g., apartment buildings).

\begin{figure*}[]
    \caption{Examples of visual data insights advocates found useful for organizing workers\label{fig:visualizations}}
    \begin{subfigure}[t]{0.4\textwidth}
        \centering
        \caption{Locations workers spent time}
        \includegraphics[height=5cm]{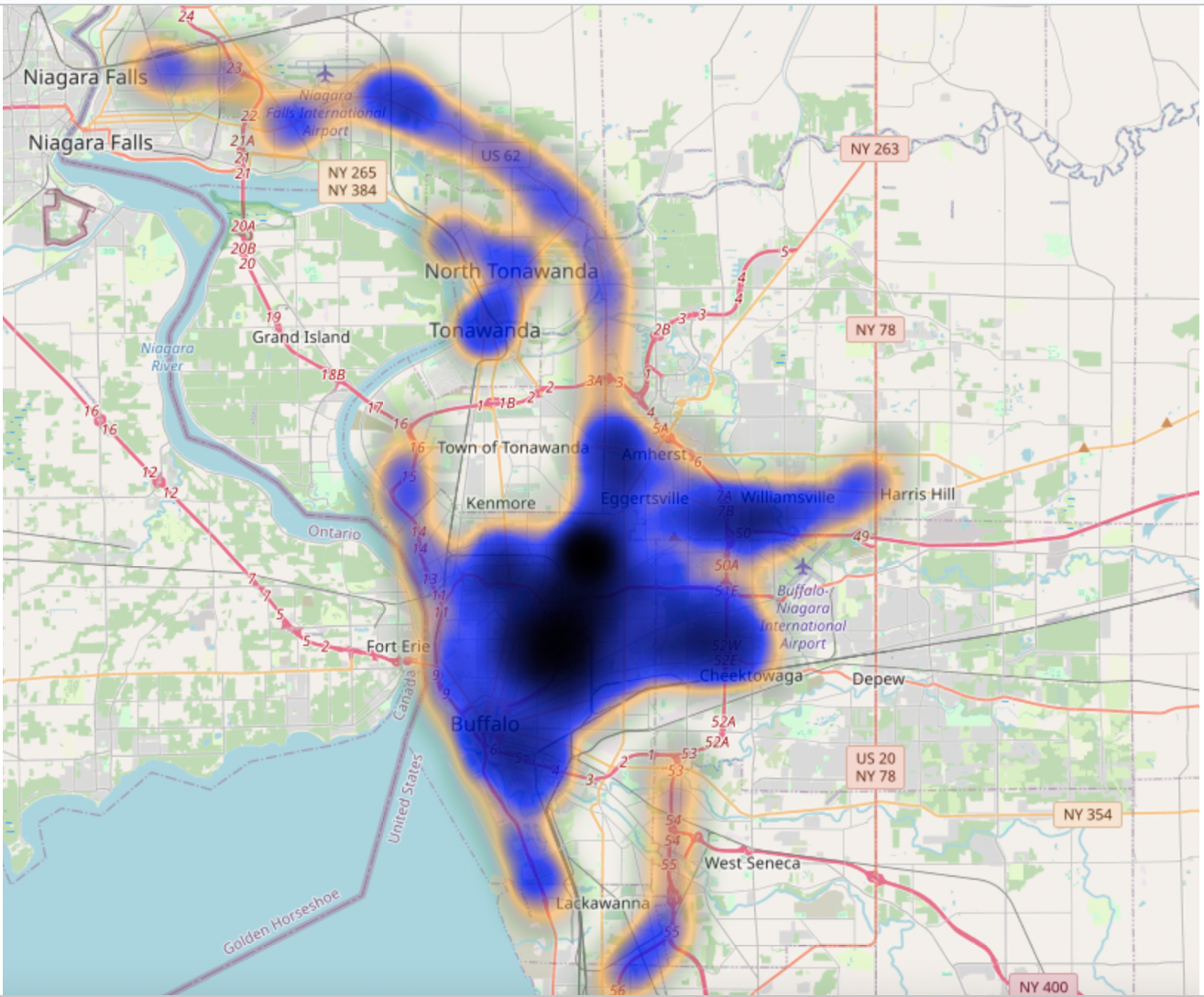} 
        \label{fig:heatmap}
    \end{subfigure}
    \begin{subfigure}[t]{0.4\textwidth}
        \centering
        \caption{Times workers work each week}
        \includegraphics[height=5cm]{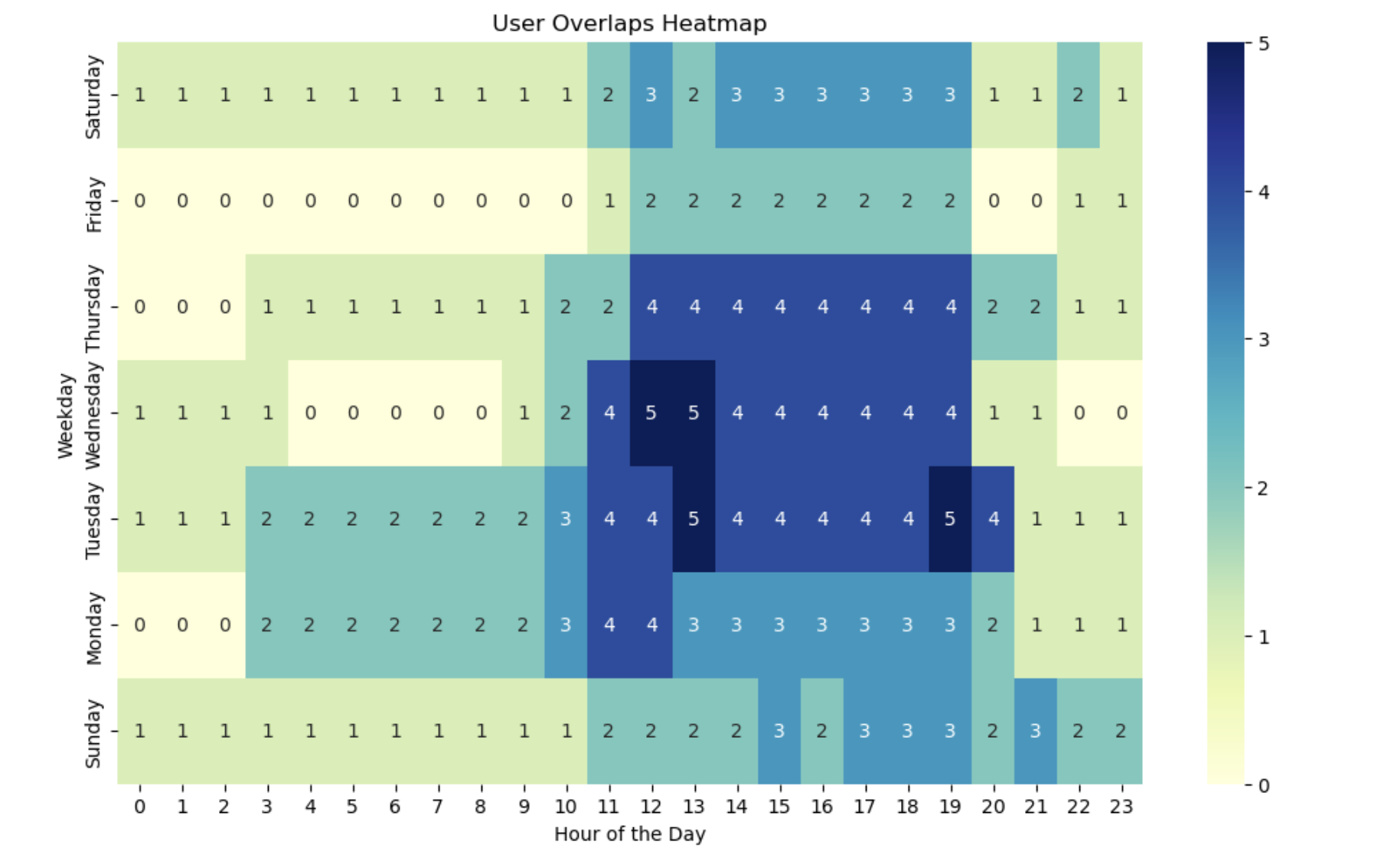} 
        \label{fig:schedule}
    \end{subfigure}
    \Description[Two types of heatmaps that were helpful for organizing workers.]{The first is a location heatmap representing the most frequently visited locations across multiple workers, with color gradients ranging from black to blue to orange indicating the locations where most workers have visited at least once. The heatmap displays the number of unique workers overlapping at each hour of each weekday, by counting workers who have work shifts during that hour. The y-axis represents days of the week, and the x-axis shows the hours of the day starting from midnight (0). On the left, there is a legend that shows how the color gradient represents the number of overlaps. The darker the shade on the heatmap, the more overlaps occur. On most days, 2-5 workers have overlapping shifts starting around mid-day and extending into the afternoon. Tuesdays and Wednesdays have the most overlap, with up to 5 workers having shifts around 11 am to 12 pm.}
\end{figure*}

\paragraph{Experiences}
The intervention did succeed in collecting data that confirmed what workers and advocates knew, but was less proficient at raising new issues or trends.
Many reactions from advocates and workers were along the lines of the data being ``\textit{not surprising or shocking to me}'' (W7) or ``\textit{exactly what I expected}'' (A6). However, advocates found that the data was pointing them in the right direction and suggested more analyses that could lead to novel insights. A6 and A14 noted that one of their current target locations of interest for worker outreach was also highlighted in our data and requested time to dive into specific times and locations where workers spent the their time\update{, where the places where more workers spent time are darker in} Figure~\ref{fig:heatmap}. The advocates also found the aggregation of worker schedules helpful to know when to target them\update{, where overlaps in hours worked are darker sections in} Figure~\ref{fig:schedule}. 

Advocates also said that even if the data did not necessarily raise new insights, it was still ``\textit{helpful to see \dots the back end of it}'' (A6) and broader trends in worker issues. 
Moreover, A14 found having the records of the issues ``\textit{does give it a degree of legitimacy}.'' Ultimately, however, some of the advocates said that they actually wanted workers not just to report the issues, but to focus on moving towards collective action:
\begin{quote}
    ``\textit{The school of thought that I come from when it comes to organizing is that appealing to the boss's humanity pretty much never works, pretty much never makes a difference. \dots It's a lot more helpful to be able to say \dots here's how we've moved these workers into action---be it a picket, be it a strike. My concern when we really get into some of this narrative end of things is that it makes \dots workers feel like `I've already done what I had to do because I reported these things' versus `I need to take action to change these things.'}'' (A14)
\end{quote}
To A14, ``\textit{aggregate information [was] definitely the most helpful and most exciting}'' to motivate collective action, rather than having workers individually address their issues.

As the advocates aimed to use the data to drive collective action, they were interested in getting more data for specific work circumstances.
Many advocates noted the ``\textit{range of issues that home care workers may be dealing with,}'' as A5 said:
\begin{quote}
    ``\textit{A lot of people think it's just taking care of older adults and the needs that come with that. We've had plenty of instances of people taking care of really, really young children, middle-aged people, people with traumatic brain injuries, and all different types of health issues that home care workers are exposed to and have to kind of navigate.}'' (A5)
\end{quote}
Our intervention was able to get diverse information and handle varied logistics such as different pay calculations for different numbers of clients (i.e., W11 had 6 clients and W7 had only two, one of which was admitted to the hospital) or ways of getting and reporting hours (i.e., the 11 workers worked for 8 different agencies). The advocates also noted the importance of the intervention collecting data about workers from Upstate New York, as often statistics about workers in New York State are skewed towards the more populous New York City. 

However, the advocates wanted even more diverse perspectives.
A1 saw that the workers who participated in the study were those who were ``\textit{more dedicated}'' or ``\textit{pretty solidly committed}'' to activism to improve their working conditions. She noted the inherent challenge to finding and collecting data from workers who may be more disgruntled or were ready to leave the job. Additionally, the model of home care that is consumer-directed, or where the client is able to pick and manage their own care workers, rather than through agencies ``\textit{is really dismissed}'' (A13) and A13 wanted more information about home care in that context specifically.

\subsubsection{Creating More Caring Policies}
\label{sec:policy}

\sparagraph{Expectations}
Advocates hoped that the data collected could help them advocate for new practices and regulations that reflect the realities of care. A6 wanted the data to help ``\textit{show a politician or employer what these people are going through}.'' A3 mentioned the importance of making home care workers and their work more visible:
\begin{quote}
    ``\textit{I've been working with home care workers for the past three years. And throughout the rest of my career, I really haven't dealt with them and didn't know much about what they did and how they did it. Just fact of how many clients are out there and how many home care workers are out there, running around, seeing four and five clients a day just to try to get full-time hours, I think that's something that can be discussed to try to persuade policymakers and let the public know that we're out there. There's a whole other community of healthcare workers that nobody sees or hears.}'' (A3)
\end{quote}
A7 similarly felt that the general public considers home care workers as ``\textit{out of sight, out of mind because it's the tough job that they [the general public] don't want to do.}''
A8, also optimistic about the power of data, noted that the ``\textit{creation of statistics can be really powerful because it's something translatable to the broader public}.'' 

The advocates found focusing on the shortage of caregivers was one way to make the plight of workers salient to the public. This shortage has been exacerbated by poor working conditions. As W11 said, ``\textit{until I started doing this job, I always wondered why  we can never keep [home care] aides, because it's a rough job \dots and not everybody is willing to do what I do}.'' As a home care recipient herself, A9 discussed that, from the client perspective, this complicates their lives because ``\textit{a lot of aides are not ideal employees because wages are so low}.'' 

One way advocates envisioned the data addressing the shortage of home care workers is making sure workers are matched well with clients, so that the demand for care is met and that workers are not overburdened in trying to meet that demand. A11 described how the current system means that ``\textit{there's lots of clients who want more hours \dots and don't get it}'' and that data about where workers and clients were located could help elucidate ``\textit{how large the care gap is for particular regions}.'' However, A1 said that ``\textit{if we could improve the matching process, and the way workers acquire hours, we would see less turnover and better outcomes for people who need care}.'' W8 noted that one option could be to have stipends for clients who are located in more rural areas to reward workers who were willing to make the trek.

Having a clearer picture of the demand and supply of home care would also help ensure more equitable distribution of money, a big concern for the advocates. The advocates discussed recent budget cuts to home care and highlighted how data on the distribution of workers and hours could help demonstrate the critical need for funding. 
A1 discussed how the cuts should not be aimed at the workers; rather, they should be aimed at the home care agencies that are ``\textit{money-grubbing, for-profit entities}.'' Moreover, advocates discussed how spending on home care could lead to public health savings in the long run because, as A2 put it, they ``\textit{heard from people where if their home care worker doesn't show up for whatever reason, they have to call the paramedics}'' and better data could help us determine ``\textit{how much money is spent by us as a society when care isn't given by home care workers}.'' A8 and A9 also posed similar questions around nursing homes as more expensive alternatives for home care.

\paragraph{Experiences}
The advocates saw the data as promising for telling stories about worker realities to the broader public, but had different ideas of how data can be used to achieve this. Many advocates thought that data was integral to advocacy work, even if they themselves were not proficient with it:
\begin{quote}
    ``\textit{At this point, we see that computers and data is what's really taken over the world. And if this is the way to reach an employer or a politician, \dots let's use these things to our advantage. \dots I don't understand half of [what] we have to do on a computer, but, this is what we have to do}.'' (A6)
\end{quote}

A14 thought policymakers and employers could be swayed by ``\textit{hard numbers and demonstrable outcomes}.'' 
However, advocates noted that the intervention would need to be scaled up to cover more home care workers to achieve this. A1 pointed out that smaller agencies have 250-400 workers, while larger ones have nearly 3,000, suggesting that data from hundreds of workers is needed to adequately represent the Upstate New York region.

The advocates also had different ideas of how the data could be best leveraged to initiate change. 
For example, advocates from HWR, with a focus on worker organizing, were more interested in worker protections. In contrast, those from NYCM, who were more involved in disability advocacy, prioritized issues related to worker distribution and care requirements. Moreover, NYCM's interest in the data was reflected by the limitations of the operations of their organization and they were most interested in the data insomuch as it would fit their priorities:
\begin{quote}
    ``\textit{Because we're such a small team, we're pretty narrow in our focus. Right now, our campaign is, focused really hard on the managed long term care, insurance companies, and getting them out of our Medicaid home care delivery system.}'' (A12)
\end{quote}

\section{Discussion}
Through designing and piloting an intervention for data-driven advocacy, we explored how we might collect quantitative and qualitative data in a way that mitigates the harms to workers (Section~\ref{sec:data-collection}) and explored how this data can be used to achieve advocacy goals (Section~\ref{sec:data-uses}), in the new context of home care work. We generated novel findings related to how workers and advocates expected and experienced data-driven advocacy, confirming, contradicting, and extending prior literature. For example, our findings elaborate on existing concepts, such as exploring ways to combine numeric and narrative data to better contextualize worker experiences and provide more power. Additionally, our findings present alternatives to what the literature or advocates expected, such as the workers being more willing to share their own data with employers but concerned about client data.

We now discuss some of the ways our findings extend prior literature \update{and present implications for data-driven advocacy in home care contexts in addition to other care and low-wage work contexts}. We \update{build on} the \update{broader} theoretical conversation \update{about data advocacy beyond human-computer interaction about} tradeoffs that should be considered around data-driven advocacy, including how to reconcile the priorities of the individual and the collective (Section~\ref{sec:individual-collective}) and the discrepancies between ideals and realities in data-driven advocacy (Section~\ref{sec:expectations-experiences}). 
We also \update{draw on human-computer interaction literature on other empirical studies of care and data to} discuss actions like combining numbers and stories (Section~\ref{sec:numeric-narrative}), leveraging advocates as stewards (Section~\ref{sec:advocates-stewards}), and tailoring towards specific goals (Section~\ref{sec:tailoring}).
\update{These tradeoffs and strategies present future work that not only further explores the theoretical expectations behind them but also the empirical experiences in different settings where low-wage, frontline workers are increasingly seeing their labor quantified (e.g., transportation \cite{Akridge2024-uj}, hospitality \cite{Spektor2023-oe}, gig work \cite{Zhang2022-zk}).}

\subsection{Additional Tradeoffs of Data-Driven Advocacy}

\subsubsection{Balancing Individual and Collective Power}
\label{sec:individual-collective}

One direct tension our work revealed is between short-term, individual burden versus collective, long-term benefit. \update{As our study found, individual workers often faced challenges in consistently engaging with the intervention given the physical and logistical infrastructure. This was compounded by the lack of a more immediate feedback loop for the workers---only two of the workers ended up finding any paycheck issues within the study period and, realistically, many would have to continue to track wage and work information without knowing if it would lead to identifying an issue. Moreover, as the advocates discussed, even if an}
employer \update{was found} violating wage policy, bringing a class or collective action case to court could take years before a worker might receive recompense. 
\update{Moreover, a}ny changes to wage or work policy for home care workers in New York State would similarly take time to enact. In the face of such long delays, what can be done to keep workers motivated towards providing data to achieve advocacy goals? 

Technological design could be used to shorten the feedback loop to demonstrate to workers that what they were doing could lead to change, \update{but simply having technological affordances is sometimes not enough.} 
A1 and A8 suggested \update{features within the technology to shorten the feedback loop} to help workers understand how their data could improve self-care or lead to further outcomes.
\update{However, as A14 noted, i}f \update{the feedback loop was closed such that} workers were nudged and rewarded for simply reporting issues, they may not want to continue with the necessary follow-up actions of contention such as protesting or going on strike. 
This finding intersects with literature on individual versus structural actions towards social change, including \update{\citet{Salehi2015-jw}'s discussion on the ``human infrastructure'' necessary to overcome challenges in technology-mediated collective action and} \citet{Brynjarsdottir2012-bm}'s discussion on how technologies that focus on persuading individuals to change their behavior often fall short of addressing structural changes required across corporations and collective attitudes.

\update{Therefore, as t}he advocates in our study noted \update{any data-driven intervention needs} to be able to contribute to all aspects of the maxim of ``educate, agitate, and organize''---helping workers reflect on their own circumstances, encouraging them towards being able to take action about it, and bringing workers together to act.
\update{This would require a careful consideration of how the intervention contributes to each of these goals in addition to how to balance what \citet{Tseng2024-fn} describe as ``benefit'' and ``burden'' in participatory systems. For example, the intervention should include features to nudge workers to engage by not only diligently tracking information to reflect on their present individual circumstances, but also contribute to collective efforts for future changes.}

\subsubsection{Integrating Expectations and Experiences of Sociotechnical Infrastructure}
\label{sec:expectations-experiences}

Another area of tension is around the distance between the hopes workers and advocates had for data-driven worker advocacy and the realities of what it looked like in practice.
\citet{Sengers2021-iu} raise a similar tension in speculative design, specifically around ``leveraging fantasy for pragmatic ends, grounding audacious fictions in imported realities.'' 
Our study is motivated by these ideals in efforts to develop ``real, existing programs of action that are intended to actualize them,'' the futures where workers and advocates could use data to contest the power employers have over home care workers.
Additionally, previous scholarship indicates that instances of fictions and realities reveal infrastructures that are ``by definition invisible \dots [but] become visible on breakdown'' \cite{Star1999-db}, which can help us situate the challenges and approaches to mitigating them within the longer history of data-driven advocacy.

Many participants had techno-optimistic notions of how data could change working conditions.
Participants were excited about how data could reach employers and politicians and were confident that information could be put towards better policies, even though they acknowledged their own lack of knowledge when it came to technology.
This type of viewpoint echoes literature in information and communication technologies for development that criticizes technology as a silver bullet solution to ``wicked problems'' \cite{Rittel1973-wk} in society \cite{Toyama2015-ko}. 

One direction for future data-driven advocacy interventions would be to more actively set expectations for workers and advocates, grounding their misunderstandings of technology in real experiences, whether these misaligned mental models are too pessimistic or too optimistic. For workers, this could involve helping them feel less fearful of new technologies but also more critical about how the inputs of their data could be used. In our study, we saw how the workers updated their beliefs throughout the study. Their attitudes toward technology evolved as they used the intervention, which could help them less immediately reject new technologies. Rather, through a more realistic understanding of how their data could be used or misused, could help them be specific about how they protect themselves.



\subsection{Future Strategies for Data-Driven Worker Advocacy}

\subsubsection{Combining Numeric and Narrative Data}
\label{sec:numeric-narrative}

\update{The advocates noted that a combination of numeric and narrative data could potentially lend the data more power in the eyes of policymakers and the public.}
\update{Prior r}esearch has \update{also} noted the importance of having both numbers and stories to represent an issue. This has been framed as ``data-plus'' \cite{Spektor2024-mo} or data that goes ``beyond the strictly representative and quantitative'' and towards ``the affective and the narrative'' \cite{Crooks2021-gw}. \update{As prior literature has noted, this is especially important for care work contexts, where it is not only difficult to capture all of the relational and emotional contributions of care \cite{Saxena2023-lj}, but also important to approach the data collection and use process with care \cite{Nielsen2023-lp}. This will also extend throughout the lifecycle of the data that is being used, as \citet{Karusala2019-do} notes, ``data may become an artifact to additionally care for, as well as a conduit for care that may or may not transmit the intended meaning.''}

\update{In our study, w}e explored how to pair quantitative with qualitative data, \update{both asking workers to track both passive background metrics and active input and} trying out computational approaches to analyzing the stories presented. \update{Through this approach, we noted the importance of}
\update{\textit{care}-}fully consider\update{ing} the extra effort of collecting and context switching between multiple forms of data, potentially by more tightly integrating different types of data. In our study, this could have looked like tying the journal entries to the GPS measurements for workers to have a chance to respond to their data and contextualize it in a way that does not require a separate process.
\update{Or we could incorporate a ``computational narrative analysis'' like \citet{Saxena2023-lj} that can interpret quantitative location data to support workers' qualitative journal responses.}
\update{This type of approach could be relevant for other geographically distributed workplaces, such as delivery drivers \cite{Dalal2023-wo}, who are also independent contractors with non-public workplaces.}


\subsubsection{Leveraging Advocates as Stewards}
\label{sec:advocates-stewards}

\update{Another way our study considered worker burden with care is that} we worked closely with a variety of advocates to alleviate some of the ``epistemic burden'' \cite{Pierre2021-gb} on workers and design the intervention to more closely align with real-world advocacy goals. Another way that working with advocates might prove helpful is to leverage their close and frequent interactions with workers to enable advocates to serve as stewards, or proxies for workers in making decisions on which data is shared and how to reduce the burden on individual workers---with the caveat that advocates need to be cautious of how their own priorities may differ from those of the workers.

For example, in relation to data privacy, we found that individual workers did not, for the most part, have strong stances on the privacy of their own data, especially in relation to it being used for broader collective advocacy. \update{Workers felt burdened rather than empowered when asked to decide which data to share and when.} Therefore, rather than asking individual workers to make decisions about how their data is used, as is described in idealized participatory privacy methods \cite{Mir2021-xo}, it may make sense to adopt an ``active steward \dots to help make decisions about their data [and] take an active role in managing this information'' \cite{Tseng2024-fn} \update{or ``shifting the burden to specialists'' \cite{Karusala2019-do}}. \update{However, the advocates also mentioned how they still had to try understand workers' perspectives and lives. Thus, future studies could continue to determine the best balance of power and burden workers perceive when making choices about questions around privacy, and optimal methods and frequency with which stewards intervene.}

\subsubsection{Tailoring Data Collection to Specific Goals}
\label{sec:tailoring}

Another result of having a range of perspectives from advocates who engage with such a heterogeneous workforce is that there were a variety of demands on whose data should be collected and for what purpose. One way to handle different asks is to develop a lightweight and customizable process that different organizations could deploy as needed. For example, as A12 mentioned, their organization was focusing on a very specific policy campaign and could only use data related to overtime cuts and their impacts on workers. Therefore, it might be helpful \update{to develop a ``adaptive'' \cite{Gajos2006-je} intervention that would allow them to} collect \update{only} the \update{sufficient} data, reducing the amount of tracking the workers would have to do and the length of time the intervention needs to be deployed. \update{In the future, this could emerge as a platform or set of tools that could be deployed in many different contexts, such as Open Data Kit \cite{Hartung2010-zd}.}

On the other hand, it could also mean aggregating even larger amounts of data and giving the option of which data to analyze for specific purposes. This could look like having a large-scale deployment of a data-driven intervention that collects a wide range of data. To account for potential worker burden, this type of approach would need to either be minimally invasive or deployed with ``measurement bursts'' of intense data collection interspersed with periods without reporting requirements \cite{Fisher2012-ma}. This type of approach would also need to account for a wide range of worker circumstances, from the variation of client circumstances and conditions to different types of hiring and payment, which means the scale will likely need to be larger.

\section{Conclusion}
In this paper, we partnered with a union-affiliated organization to design and pilot intervention that supports data-driven advocacy for home care workers. We examined the expectations and experiences of advocates and workers about how data could be collected and used for data-driven advocacy. 
We saw both workers and advocates mentioning the possibility to change narratives of care, using data as both a diagnostic for problems and evidence of violations. On the other hand, we saw participants mentioning that the process could also result in increased effort and privacy risks for workers. However, through our pilot deployment, we found that workers were less concerned about their own privacy and developed more confidence in their use of the intervention over time. We also found indicators that qualitative data about the more invisible aspects of work could inform organizing strategy as well as policymakers and the broader public, but required larger scale data collection for diverse types of workers. Our findings present tradeoffs in individual and collective power as well as in the ideals and realities of data-driven advocacy. We discuss new strategies and considerations for future data-driven advocacy around what data to collect, the role of partner organizations, and setting expectations. Our empirical findings about data-driven advocacy could benefit other low-wage, frontline health workers like home care workers who are a historically marginalized and spatially isolated workforce.

\begin{acks}
We would like to thank the contributions of the home care workers we engaged with and the partner organizations that helped make these connections possible. We would also like to thank the WeClock team for supporting our efforts in adapting the application to the home healthcare context, especially Dr. Dana Calacci. This research was funded by a Google Cyber NYC grant.
\end{acks}

\bibliographystyle{ACM-Reference-Format}
\bibliography{paperpile,bib}

\end{document}